\documentclass[journal]{IEEEtran}

\usepackage{amssymb}
\usepackage{amsmath}
\usepackage{amsfonts}
\usepackage{multirow}
\usepackage{graphicx}
\usepackage{textcomp}
\usepackage{amsthm}
\usepackage{setspace}
\usepackage{cite}

\usepackage[noend]{algpseudocode}
\usepackage{algorithmicx,algorithm}
\usepackage{subfigure}
\usepackage{multicol}
\usepackage{color}
\usepackage{xspace}
\usepackage{tabularx}
\usepackage{array}
\usepackage{makecell}
\usepackage{bm}
\usepackage[marginal]{footmisc}

\begin{document}

\title{Adaptive Multi-Feature Budgeted Pro\mbox{f}it Maximization in Social Networks}

\author{Tiantian~Chen,
        Jianxiong~Guo,~\IEEEmembership{Member,~IEEE}
        Weili Wu,~\IEEEmembership{Senior Member,~IEEE} % <-this % stops a space
\thanks{T. Chen and W. Wu are with the Department of Computer Science, Erik Jonsson School of Engineering and Computer Science, University of Texas at Dallas, Richardson, TX 75080, USA. \textit{(Corresponding author: Tiantian Chen.)}}% <-this % stops a space
\thanks{E-mail: tiantian.chen@utdallas.edu.}
\thanks{J. Guo is with the BNU-UIC Institute of Artificial Intelligence and Future Networks, Beijing Normal University at Zhuhai, Zhuhai, Guangdong 519087, China, and also with the Guangdong Key Lab of AI and Multi-Modal Data Processing, BNU-HKBU United International College, Zhuhai, Guangdong 519087, China.}
}

%\thanks{Manuscript received XXX; revised XXX.}}


\maketitle

\newtheorem{theorem}{Theorem}
\newtheorem{lemma}{Lemma}
\newtheorem{corollary}{Corollary}
\newtheorem{definition}{Definition}
\newtheorem{proposition}{Proposition}
\newtheorem{remark}{Remark}
\newtheorem{claim}{Claim}
\renewcommand{\algorithmicrequire}{\textbf{Input:}}
\renewcommand{\algorithmicensure}{\textbf{Output:}}

\begin{abstract}
Online social network has been one of the most important platforms for viral marketing. Most of existing researches about diffusion of adoptions of new products on networks are about one diffusion. That is, only one piece of information about the product is spread on the network. However, in fact, one product may have multiple features and the information about different features may spread independently in social network. When a user would like to purchase the product, he would consider all of the features of the product comprehensively not just consider one.
%There are indeed some works about multiple diffusions. However, they are either about competitive products or about complementary products, not about only one product.
Based on this, we propose a novel problem, multi-feature budgeted pro\mbox{f}it maximization (MBPM) problem, which first considers budgeted pro\mbox{f}it maximization under multiple features propagation of one product. 

Given a social network with each node having an activation cost and a profit, MBPM problem seeks for a seed set with expected cost no more than the budget to make the total expected pro\mbox{f}it as large as possible. We mainly consider MBPM problem under the adaptive setting, where seeds are chosen iteratively and next seed is selected according to current diffusion results. We study adaptive MBPM problem under two models, oracle model and noise model. The oracle model assumes conditional expected marginal pro\mbox{f}it of any node could be obtained in $O(1)$ time and a $(1-1/e)$ expected approximation policy is proposed. Under the noise model, we estimate conditional expected marginal pro\mbox{f}it of a node by modifying the EPIC algorithm and propose an e\mbox{ff}icient policy, which could achieve a $(1-e^{-(1-\epsilon)})$ expected approximation ratio. Several experiments are conducted on six realistic datasets to compare our proposed policies with their corresponding non-adaptive algorithms and some heuristic adaptive policies. Experimental results  show e\mbox{ff}iciencies and superiorities of our policies.

\end{abstract}


\IEEEpeerreviewmaketitle

\section{Introduction}
Online social network, like Facebook, Twitter, Linkedln, etc., has been one of the most important platforms for marketing and communication. Many companies have taken social network as main method to promote products by word-of-mouth effects. To maximize the product influence and obtained pro\mbox{f}it, companies may apply many methods, such as distributing coupons, free samples or offering some discounts when purchasing. Many researches have been focused on the diffusion phenomenon on social networks, including diffusion of ideas, news, adoptions of new products, etc. One topic extensively studied is the Influence Maximization (IM) problem \cite{Borodin,Chen09,Chen101,Chen102,Kempe03}, which asks for $k$ seeds to maximize the expected number of influenced users under some diffusion model. There are two classical diffusion models: Independent Cascade (IC) model and Linear Threshold (LT) model. 

However, it has been proved IM problem is NP-hard and computing the expected spread from a node set is \#P-hard in general under IC and LT model \cite{Chen101, Chen102}. Kempe \textit{et al.} \cite{Kempe03} presented a $(1-1/e)$-approximation scheme, classical greedy algorithm, for IM problem and they used Monte Carlo method to estimate expected spread of a seed set, but it is time-consuming. Many recent works, like \cite{Tang14, Tang15, Nguyen171, Huang, Han,Huang2,tang2018online}, have been focused on solving this problem, which not only could obtain a $(1-1/e-\epsilon)$-approximation solution with high probability but are efficient even for large-scale datasets. 

Most of existing papers related to IM problem only consider a single diffusion. That is, only one piece of information about the product is spread on social networks. Some papers  like \cite{Chen, Guo19, Zhang, Zhangy} indeed consider multiple diffusions of products. But the diffusions are for multiple products and each diffusion is for one product. However, in reality, one product may have multiple features and the information about all these features can spread on social network. For instance, when a customer wants to buy a phone, he may consider many features, such as price, brand, camera, display, speed, etc. He has his own preference for each feature, which can be regarded as weight for the feature, and has a threshold to purchase the phone. He heard the information about features of the phone on networks, and he will purchase the phone only when the sum of weights of features satisfying his requests is larger than or equal to  the threshold.

Guo \textit{et al.} \cite{Guo2} first proposed a multi-feature diffusion model (MF-model) to describe multiple features about one product spreading on the social network, where different feature information spreads independently according to different successful probabilities and whether to accept a product is determined by overall evaluation on all these features. They considered the rumor blocking problem under this model, but they assume the weights of each feature for each node are equal when solving the problem.

Based on the MF-model \cite{Guo2}, we propose a novel budgeted profit maximization problem, MBPM problem. Given a social network, MBPM problem assumes that multiple information about multiple features of a product are spread on it. Each feature has its own propagation probability when spreading from one user to another, and each user has its own weights for each feature. A user will purchase the product only when the sum of weights of features he accepts is larger than or equal to his threshold.  Each node has an activation cost and a profit. MBPM problem seeks for a seed set with expected cost no more than the budget to make the obtained profit as large as possible. We consider MBPM problem under the adaptive setting, in which next seed is selected based on the diffusion result of current seeds. That is, we first select a seed and then observe which nodes would be activated by the seed. According to diffusion result, we would select next seed to maximize the profit as much as possible.
 
Main contributions of this work are as follows:
%\vspace{-0.1in}
\begin{itemize}
\setlength{\itemsep}{0pt}
\setlength{\parsep}{0pt}
\setlength{\parskip}{0pt}
       \item  We propose a novel practical problem, MBPM problem, consider it under both the non-adaptive and adaptive settings and propose efficient strategies to solve them, respectively.
       \item  For the non-adaptive MBPM problem, we show its objective is monotone submodular and give a randomized algorithm which could achieve $(1-1/e)$ expected approximation guarantee.
       \item For the adaptive MBPM problem, we prove its objective function is adaptive monotone and adaptive submodular. We consider adaptive MBPM problem under two models, oracle model and noise model. An policy with $(1-1/e)$ expected approximation ratio is given in oracle model. Under noise model, we estimate the conditional expected marginal pro\mbox{f}it of any node by modifying the EPIC algorithm and propose a sampled adaptive greedy policy which could achieve a $(1-e^{-(1-\epsilon)})$ expected approximation guarantee, where $0<\epsilon<1$.
       \item Experimental results on realistic datasets confirm effectiveness and superiority of our algorithm.
\end{itemize} 
\vspace{-0.05in}

\medskip

\noindent \textbf{Organization.} In Section \ref{related work}, we introduce related works of MBPM problem. The multi-feature diffusion model (MF-model) and IM problem under the MF-model are discussed in Section \ref{problem formulation}. Section \ref{MBPM} presents the MBPM problem under the non-adaptive setting and its solving algorithm. Section \ref{AMBPM} gives de\mbox{f}inition of adaptive MBPM problem and property of its objective function. Section \ref{algorithm} gives the algorithm to solve the adaptive MBPM problem and corresponding proofs for theoretical guarantee. Section \ref{Experiments} is dedicated to experiments and Section \ref{Conclusion} concludes the paper.

\section{Related Works}\label{related work} 
Kempe \textit{et al.} \cite{Kempe03} first formulated IM problem as a combinatorial optimization problem, which aims to choose $k$ seeds to make the expected in\mbox{f}luence as large as possible. They presented a $(1-1/e)$-approximation algorithm, classical greedy scheme, to solve IM. Later, many variants of IM problem appeared, such as coupon based profit maximization \cite{Liu, Tong, Guo2020}, multiple products profit maximization \cite{Chen, Zhang, Zhangy}, etc. The one related to our work is cost-aware targeted viral marketing (CTVM) problem \cite{Nguyen172}, which maximizes the expected total benefit by choosing a seed set under the budget. The difference between CVTM and our MBPM problem is CVTM only considered one information diffusion on networks under the classical IC and LT model. And they studied CVTM problem under non-adaptive setting and designed a $(1-1/\sqrt{e}-\epsilon)$-approximation algorithm. Banerjee \textit{et al.} \cite{Banerjee} considered targeted CVTM problem where only nodes in target set have an activation profit, and they proposed a $(1-1/\sqrt{e})$-approximation algorithm. However, we consider multiple diffusions of a product's features and studied the MBPM problem under the adaptive setting. \cite{Shan} considered a diff\mbox{usion} model of multiple diffusions of a product's features, which is similar to ours, but the activation threshold for each node in their model is fixed. They measured the amount of information that a user received as the probability that the user is activated in an information cascade. However, in our MF-model, each node has a weight for each feature which measures how the user cares about the feature, and the activation threshold for each node is distributed uniformly in [0, 1].

For adaptive problem related research, Golovin \textit{et al.} \cite{Golovin} proved the objective of adaptive IM problem is adaptive monotone and submodular under full-adoption model and IC model. They proposed an adaptive greedy scheme, which is a $(1-1/e)$-approximation scheme for adaptive IM problem. They also proved this algorithm can be used when a set function with adaptive monotonicity and submodularity subjects to the knapsack constraint. Han \textit{et al.} \cite{Han} considered an variant of the adaptive IM problem, where $k$ seeds are selected in batches of equal size $b$. They designed an  AdaptGreedy framework instantiated by scalable IM algorithms, which could achieve a $(1-e^{-(1-1/e)+\epsilon})$ approximation guarantee with high probability. Sun \textit{et al.} \cite{Sun} studied a Multi-Round Influence Maximization problem under the adaptive setting where information spreads in multiple rounds independently from probably different seed sets. They proposed an adaptive algorithm instantiated by the IMM \cite{Tang15}, which could guarantee $(1-e^{-(1-1/e)}-\epsilon)$ approximation. Recently, Huang \textit{et al.} \cite{Huang2} pointed out there are some gaps in analysis of approximation guarantee for adaptive policies in \cite{Han} and \cite{Sun}. They fixed the previous AdaptGreedy framework in \cite{Han} by instantiating with their improved EPIC algorithm and showed it could provide a  $(1-e^{\rho_{b}(\epsilon-1)})$ expected approximation guarantee. \cite{Guo20} considered the adaptive influence maximization with multiple activations problem, where a selected node in each iteration can be unwilling to be the seed and a node not being the seed in previous iteration can be activated again later but with higher activation cost. The goal is to find a randomized policy subject to expected knapsack constraint to maximize the expected influence spread. They designed an adaptive greedy policies by modifying EPIC algorithm in \cite{Huang2}. Peng \textit{et al.} \cite{Peng} showed that the adaptivity gap of the IM problem under the IC model with myopic feedback is at least $e/(e-1)$ and at most 4, and that both the non-adaptive and adaptive greedy algorithms achieve a $\frac{1}{4}(1-1/e)$-approximation to the adaptive optimum. \cite{Chen19} showed the adaptivity gap of the IM problem under the IC model with the full-adoption feedback on several families of well-studied influence graphs. \cite{Chen20} proposed the concept of greedy adaptivity gap, comparing the performance of adaptive greedy algorithm to its non-adaptive counterpart.

\section{Inf\mbox{l}uence Maximization Problem under the MF-model}
\label{problem formulation}
A social network is generally denoted by a directed graph $G=(V, E)$, where $|V|=n$ and $|E|=m$. For each $(v, w)\in E$, $v$ is named the in-neighbor of $w$ and $w$ is called the out-neighbor of $v$. Here, each node $u\in V$ represents a user (customer) in this paper.

\subsection{Multi-Feature Diffusion Model}
Consider a product with multiple features and the information about each feature may spread from one customer to another. To characterize it, we consider the multi-feature diffusion model (MF-model) \cite{Guo2} in this paper.

\textrm{1.}  Given a social network $G=(V, E)$, $q$ pieces of  information about $q$ features of a product are spread on it, respectively. For each $(u, v)\in E$, there is a $q$-dimensional propagation probability vector $\bar{p}_{u, v}=(p_{u, v}^{1}, \ldots, p_{u, v}^{q})$ associated with it, where $p_{u, v}^{i}\in (0, 1]$ is the successful probability when $u$ tries to motivate $v$ to accept the information about feature $i$ of the product.

\textrm{2.} Each $u \in V$ has a threshold $\theta_{u}$ distributed in [0,1] uniformly and a weight vector $\bar{w}_{u}=(w_{u}^{1}, \ldots, w_{u}^{q})$, where $w_{u}^{i}$ denotes the weight of feature $i$ for user $u$ and $\sum_{i=1}^{q}{w_{u}^{i}}=1$.

\textrm{3.} When user $u$ accepts feature $i$ at timestamp $t$ (called $i$-accepted, otherwise called $i$-unaccepted), it will attempt to motivate its $i$-unaccepted out-neighbor $v$ with successful probability $p_{u, v}^{i}$ at timestamp $t+1$. The information about different features is diffused independently on the social network, and user $v$ will purchase the product if and only if the sum of corresponding weights of features that have already been accepted by  $v$ is no less than $\theta_{v}$ (called purchase condition). 
%When a node $v$ purchases the product, it is called active. Otherwise, it is called inactive.

\textrm{4.} Initially, a set of seeds is activated to spread all of the $q$ features. At every step, each node that hasn't purchased the product would check whether the purchase condition is satis\mbox{f}ied. The diffusion process will continue until there is no more node activated. 

To further illustrate the model by the phone example, say, the five features of a phone corresponding to price, brand, camera, display, and speed are 1, 2, 3, 4, and 5, respectively. Each node can either accept or not accept each feature. For example, a potential customer may either be convinced that the display is good or not. Each node $v\in V$ also has a weight for each of the five features: $w_{v}^{1}, w_{v}^{2}, w_{v}^{3}, w_{v}^{4},$ and $w_{v}^{5}$, which measures how much he cares about each feature. Before the diffusion, each node also has to sample a threshold $\theta_v\in [0, 1]$. Initially, all features for each seed are accepted. Then, we have five different cascade processes corresponding to the five features. Each of them follows the IC model, and the five cascade processes are independent. Now, at the end of the cascade process, each node is infected by some of the five features. A node will eventually buy the product if the sum of the weights of the accepted features exceeds its threshold. For example, if $v$ accepts features 2 and 4 but not 1, 3, 5, then $v$ will be considered activated if the weighted sum $w_{v}^{2}+w_{v}^{4}$ exceeds his threshold $\theta_{v}$.

\iffalse
Let's see the inspiration behind the assumption that information of different features diffuse independently. On the one side, a user can accept some features of the phone even though he doesn't buy it.

On the other side, a user buy the phone may not accept all the features. It's kind of not realistic for the model in which a user buying the product can activate its neighbor to accept all features. This is because that it's not convincing when you try to let other people believe the things which you don’t even believe. 
\fi

Even though Guo \textit{et al.} \cite{Guo2} first proposed this MF-model, they assume the weight of each feature for each node is the same, namely $w_{u}^{i}=w_v^i$ for any user $u, v\in V$, in their submodularity proof and algorithm analysis. This is only a special case and not that realistic. In this paper, we consider the general case where the weight vector $\bar{w}_{u}=(w_{u}^{1}, \ldots, w_{u}^{q})$ for each node $u$ is arbitrary. Denote by $\sigma(S)$ the expected number of nodes (users) purchasing the product when $S$ is the seed. Actually, we could prove that $\sigma(S)$ is monotone non-decreasing and submodular with respect to $S$ under the general MF-model. Our following analysis is based on the general MF-model, which is an important improvement and extension. Before showing properties of $\sigma(S)$, let's first see the equivalent diffusion process of the MF-model.

\begin{remark}
For convenience, we still use $G=(V, E)$ to represent the social network $G=(V, E)$ with propagation probability $p: E \rightarrow (0, 1]^{q}$, threshold $\theta: V\rightarrow [0, 1]$, and weight $w: V \rightarrow [0, 1]^{q}$. 
\end{remark}

\subsection{Equivalent Diffusion Process}
Since information of different features is spread independently on the social network in MF-model, that is, the diffusion of one piece of information about one feature has no interference on information of other features, we can view the propagation process of MF-model as follows. 

\begin{definition}[Multi-level Graph]
Given a social network $G=(V,E)$, define its multi-level graph  as $\widehat{G}=(\widehat{V}, \widehat{E})=G^{1}\cup G^{2}\cup\cdots G^{q}$, where $G^{i}=(V^{i}, E^{i})$ and each node $v_{i}\in V^{i}$ is a copy node of $v \in V$, called feature node of $v$. For each $(u, v)\in E$, there is a corresponding edge $(u_{i}, v_{i})\in E^{i}$, $i=1, \cdots, q$, and the propagation probability on $(u_{i}, v_{i})$ is $p_{u, v}^{i}$, that is, the successful probability when $u$ attempts to motivate $v$ to accept feature $i$. 
\end{definition}

\begin{figure}[htbp]
\centerline{\includegraphics[scale=0.085]{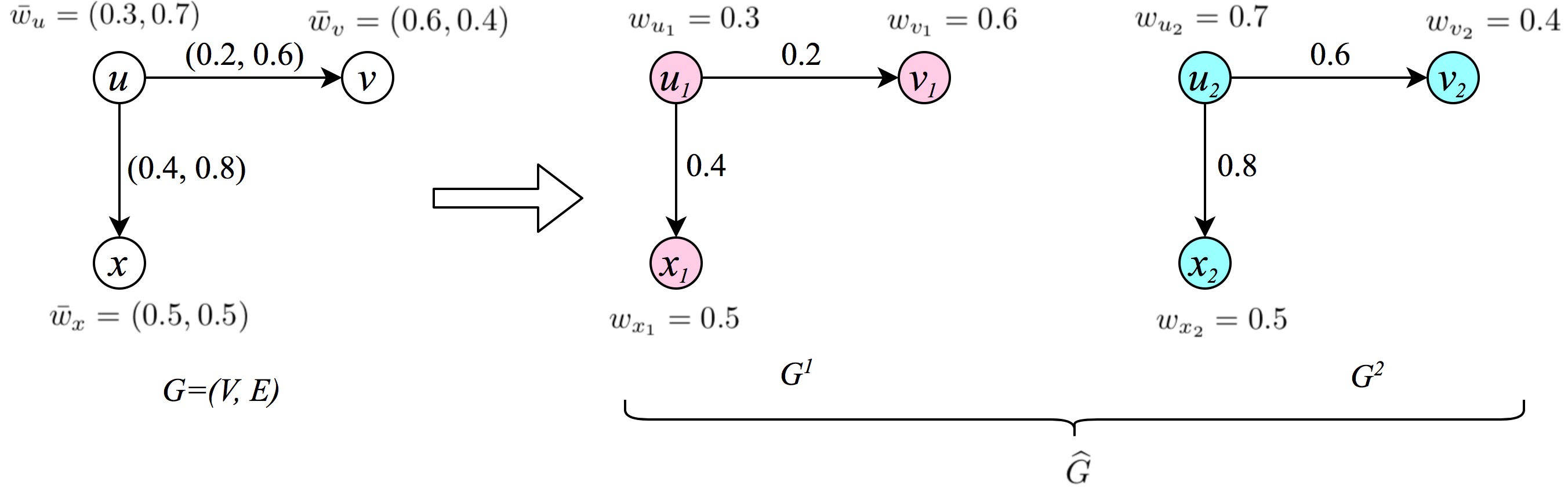}}
  \caption{An example of $G=(V, E)$ with its multi-level graph $\widehat{G}$.}
\label{multi-level-graph}
\end{figure}
%$\bar{w}_{u}=(0.3, 0.7), \bar{w}_{v}=(0.6, 0.4), \bar{w}_{x}=(0.5, 0.5)$ ${w}_{u_{1}}=0.3$\\${w}_{u_{2}}=0.7$ ${w}_{v_{1}}=0.6$\\${w}_{v_{2}}=0.4$ ${w}_{x_{1}}=0.5$\\${w}_{x_{2}}=0.5$\\
An example of the multi-level graph can be seen in Fig. \ref{multi-level-graph}. For each node set $S\subseteq V$, denote by $\widehat{S}= S^{1}\cup \ldots \cup S^{q}$ the corresponding feature node set in $\widehat{G}$ of nodes in $S$, where $S^{i}=\{u_{i}\in V^{i}: u_{i}\ \text{is the corresponding feature node of}\ u \in S \text{ for feature } i\}$. For each $u\in V$, denote the corresponding feature node set of $u$ as $\widehat{u}=\{u_{1}, \ldots, u_{q}\}$. Then we could give the equivalent diffusion process of the general MF-model.

\textrm{1.} Given a social network $G=(V, E)$ and its multi-level graph $\widehat{G}=G^{1}\cup \ldots \cup G^{q}$, $q$ pieces of information about $q$ different features are spread on $\widehat{G}$, but the information about feature $i$ is only spread on $G^{i}$. Each node $v\in V$ samples a threshold $\theta_v$ independently uniformly at random from [0,1].

\textrm{2.} Initially, we choose the seed set $S\subseteq V$ for the product. Then nodes in $\widehat{S}$ are seeds of the corresponding features.

\textrm{3.} The information about different features is diffused independently from their own seeds according to the classic IC model. A node in $G^{i}$ can only have two states:  $i$-accepted or $i$-unaccepted. A node  in $G^{i}$ accepting the information of feature $i$ is called $i$-accepted. Otherwise, it is called $i$-unaccepted.
 
\textrm{4.} After the propagation process of all features terminates, we could determine whether each node $v \in V$ would purchase the product. That is, we would check whether the sum of weights of $i$-accepted nodes $v_{i}, i=1, \cdots, q,$ is larger than or equal to $\theta_{v}$. If it satis\mbox{f}ies the purchase condition, then node $v$ would purchase the product and we call $v$ active. Otherwise, $v$ is called inactive.

\begin{remark}
To avoid confusion, we will use "infect" when we say a feature node $u_{i}$ tries to activate its out-neighbor $v_{i}$ to accept feature $i$, and use "activate" for user node. 
\end{remark}

\subsection{Property of $\sigma(S)$}

\begin{definition}[Realization]
Given a social network $G=(V, E)$ with probability $p: E\rightarrow (0, 1]$, a (full) realization $\phi$ of $G$ is defined as $\phi:E\rightarrow \{0, 1\}$. For $e\in E$, $\phi(e)=0$ (resp. 1) means edge $e$ is blocked (resp. live) under $\phi$.
\end{definition}

\iffalse
\begin{remark}
In other works, they usually call an instance of a probabilistic graph as "realization" or  "possible world". In this paper, we call it "graph realization" to distinguish with a different concept "realization", which will be introduced later.  
\end{remark}
\fi

Let $\Phi$ be a random variable denoting a random realization. Then we have
\begin{equation}
 \begin{aligned}
\Pr{[\phi]}&:=\Pr[\Phi=\phi]=\prod_{\substack{e\in E:\\  \phi(e)=1}}{p_{e}}\prod_{\substack{e\in E:\\  \phi(e)=0}}{(1-p_{e})}\nonumber\\
%&=\prod_{\substack{e\in \widehat{E}:\\  \phi(e)=1}}{p_{e}}\prod_{\substack{e\in \widehat{E}:\\  \phi(e)=0}}{(1-p_{e})}\nonumber\\
\end{aligned}
\end{equation} 

%The probability of generating a realization $g$ sampled from $G$ is $\Pr{[g]}=\prod_{e\in E(g)}{p_{e}}\prod_{e\in E\setminus E(g)}{(1-p_{e})}$.
Denote by $\Omega$ the set of all possible realizations of multi-level graph $\widehat{G}$. Let $S$ be the seed set of the product and $\widehat{S}=S^{1}\cup \ldots \cup S^{q}$ be its corresponding feature node set. Then $S^{i}$ is the seed set of feature $i$ in $G^{i}$. Given a realization $\phi\in \Omega$, for each node $u_{i}$ in $\widehat{G}$, define  
\[
x_{\phi}(S^{i}, u_{i})=
\begin{cases} 
1,  & u_{i}\ \mbox{accepts feature}\ i\ \mbox{in}\ \phi\ \mbox{under}\ S^{i}\\
0, & \mbox{otherwise}
\end{cases}
\]
Therefore,  node $u \in V$ will purchase the product under $\phi$ if and only if $\sum_{i=1}^{q}{x_{\phi}(S^{i}, u_{i})}\cdot w_{u}^{i}\geq \theta_{u}$. Denote $I_{\phi}(S^{i})$ as the node set in $G^{i}$ containing the $i$-accepted nodes in $\phi$ under $S^{i}$.That is, $I_{\phi}(S^{i})=\{u_{i}\in V^{i}|x_{\phi}(S^{i}, u_{i})=1\}$. Let $I_{\phi}(S)$ be the set of active nodes in $V$ when diffusion process of nodes in $\widehat{S}$ on $\phi$ terminates.  Then for any $u\in V$, we have  
\[
\Pr[u\in I_{\phi}(S)]=\sum_{\substack{1\leq i \leq q:\\  u_{i}\in I_{\phi}(S^{i})}}{w_{u}^{i}}.
\]
\begin{remark}
$I_{\phi}(S^{i}), i=1, \ldots, q$ are deterministic sets while $I_{\phi}(S)$ is a random set.
\end{remark}

\begin{theorem}\label{monothm}
$\sigma(S)$ is monotone non-decreasing and submodular with respect to $S$.
\end{theorem}
\begin{IEEEproof}
For any node set $S\subseteq T\subseteq V$, denote by $\hat{S}=S^{1}\cup \ldots \cup S^{q}$ and $\hat{T}=T^{1}\cup \ldots \cup T^{q}$ their corresponding copy node set in $\hat{G}$, respectively. Then $\sigma(S)$ under the MF-model can be represented as:
\begin{equation}
 \begin{aligned}
  \sigma(S)&=\sum_{\phi\in \Omega}\Pr[\phi]\cdot \sum_{v\in V}\Pr[v\in I_{\phi}(S)]\nonumber\\
  &=\sum_{\phi\in \Omega}\Pr[\phi]\cdot \sum_{v\in V}{\sum_{\substack{1\leq i \leq q:\\  v_{i}\in I_{\phi}(S^{i})}}{w_{v}^{i}}}
\end{aligned}
\end{equation} 
Since $S\subseteq T$, then $S^{i}\subseteq T^{i}, i=1, \ldots, q$. Clearly, $I_{\phi}(S^{i})\subseteq I_{\phi}(T^{i})$ since any node in $I_{\phi}(S^{i})$ can also be $i$-accepted under $T^{i}$ in $\phi$. Therefore, $\sum_{v_{i}\in I_{\phi}(S^{i})} w_{v}^{i}\leq \sum_{v_{i}\in I_{\phi}(T^{i})} w_{v}^{i}$ and $\sigma(S)$ is monotone with respect to $S$.
For any $u \in V\setminus T$,  
\[\sigma(S\cup \{u\})-\sigma(S)=\sum_{\phi\in \Omega}\Pr[\phi]\cdot \sum_{v\in V}{\sum_{\substack{1\leq i \leq q:\\  v_{i}\in (I_{\phi}(S^{i}\cup \{u_{i}\})\setminus I_{\phi}(S^{i}))}}}{w_{v}^{i}}\]
\iffalse
\begin{equation}
 \begin{aligned}
 &\sigma(S\cup \{u\})-\sigma(S)\\
=&\sum_{g\in \Omega}\Pr[g]\cdot \sum_{v\in V}{\sum_{\substack{1\leq i \leq q:\\  v_{i}\in (I_{g}(S^{i}\cup \{u_{i}\})\setminus I_{g}(S^{i}))}}}{w_{v}^{i}}\nonumber
\end{aligned}
\end{equation} 
\fi
$I_{\phi}(S^{i}\cup\{u_{i}\})\setminus I_{\phi}(S^{i})$ contains the nodes in $G^{i}$ that can only be infected by $u_{i}$ but cannot by $S^{i}$ under $\phi$. Clearly, $\left(I_{\phi}(S^{i}\cup\{u_{i}\})\setminus I_{\phi}(S^{i})\right) \supseteq \left(I_{\phi}(T^{i}\cup\{u_{i}\})\setminus I_{\phi}(T^{i})\right)$, since $u_{i}$ could infect more feature nodes when adding to $S^{i}$ than $T^{i}$ under $\phi$.  
Therefore, $\sigma(S\cup \{u\})-\sigma(S)\geq \sigma(T\cup \{u\})-\sigma(T)$ and the proof of Theorem $\ref{monothm}$ is completed.
\end{IEEEproof}

\section{Multi-feature Budgeted Profit  Maximization Problem}\label{MBPM}
A company wants to promote a new product by distributing coupons on social networks to maximize its pro\mbox{f}it as much as possible. However, the advertisement budget is usually limited. Thus, it is important to wisely select customers to allocate coupons. The product has multiple features and the information about each feature spreads from one customer to another. Given the social network $G=(V, E)$, for each $u\in V$, assume the cost of picking $u$ as the seed of product and profit obtained when $u$ purchases the product are $c(u)$ and $b(u)$, respectively. For any $S\subseteq V$, the activation cost and profit of $S$ are defined as $c(S)=\sum_{u\in S}{c(u)}$ and $\sum_{u \in S}{b(u)}$, respectively. Since we will consider randomized algorithm in the adaptive case later,  so in this section we will also consider the randomized algorithm for comparison. Given a budget $B$, we want to find a seed set $S$ with expected cost at most $B$ to maximize the obtained profit.

%Denote $\widehat{G}=(\widehat{V}, \widehat{E})$ as the social network $G=(V, E)$ with propagation probability $p: E \rightarrow [0, 1]^{q}$, threshold $\theta: V\rightarrow [0, 1]$ and weight $w: V \rightarrow [0, 1]^{q}$.

 \subsection{Problem Definition}

\begin{definition}[Multi-feature Budgeted Profit  Maximization (MBPM) Problem]
 Given $G=(V, E)$, $q$ pieces of information about $q$ features of a product are spread on the social network according to the MF-model.  The MBPM problem seeks for a seed set $S\subseteq V$ with expected activation cost at most $B$, i.e., $\mathbb{E}[c(S)]\leq B$, to maximize the total expected profit $P(S)$.
 \end{definition}

%For each node $u\in \widehat{V}$, assume the cost of activating $u$ as a seed of the product and the profit obtained when $u$ purchases the product are $c(u)$ and $b(u)$, respectively. Now we can give the equivalent diffusion process of the MBPM problem. 

Given the  equivalent diffusion process of the MF-model, we could solve the MBPM problem by solving the profit maximization problem on the multi-level graph.

Then the MBPM problem can be formulated as:
\begin{equation}
 \begin{aligned}
  & \max\ \,\, \ P(S)=\sum_{\phi\in \Omega}\Pr[\phi]\cdot \sum_{u\in V}\Pr[u\in I_{\phi}(S)]\cdot {b(u)}\\
  &s.t. \quad \mathbb{E}[c(S)]\leq B 
\end{aligned}
\end{equation} 
Based on the proof of Theorem \ref{monothm}, we have the following result.
\begin{theorem}\label{thm1}
The objective function of MBPM Problem is monotone submodular with respect to the seed set of the product. 
\end{theorem}
\iffalse
\begin{IEEEproof}
For any set $S\subseteq T\subseteq V$,  denote their corresponding node set in $\widehat{G}$ as $\widehat{S}= S^{1}\cup \ldots \cup S^{q}$ and $\widehat{T}= T^{1}\cup \ldots \cup T^{q}$, respectively. Then we have
\begin{equation}
 \begin{aligned}
  P(S)&=\sum_{g\in \Omega}\Pr[g]\cdot \sum_{u\in V}\Pr[u\in I_{g}(S)]\cdot {b(u)}\nonumber\\
         &=\sum_{g\in \Omega}\Pr[g]\cdot \sum_{u\in V}[\sum_{\substack{1\leq i \leq q:\\  u_{i}\in I_{g}(S^{i})}}{w_{u}^{i}}]\cdot {b(u)}\\
         &=\sum_{g\in \Omega}\Pr[g]\cdot \sum_{u\in V}{b(u)} \cdot \sum_{\substack{1\leq i \leq q:\\  u_{i}\in I_{g}(S^{i})}}{w_{u}^{i}}
\end{aligned}
\end{equation} 
Since we have shown $\sum_{v_{i}\in I_{g}(S^{i})} w_{v}^{i}\leq \sum_{v_{i}\in I_{g}(T^{i})} w_{v}^{i}$ and $\left(I_{g}(S^{i}\cup\{u_{i}\})\setminus I_{g}(S^{i})\right) \supseteq \left(I_{g}(T^{i}\cup\{u_{i}\})\setminus I_{g}(T^{i})\right)$, it is clear $P(S)$ is monotone and submodular with respect to $S$.
\end{IEEEproof}
\fi

\subsection{Algorithm}
%\textcolor{red}{When both the activation cost and profit of each node are 1 and $q=1$, the MPM problem aims to find a size-$k$ seed set to maximize the number of activated nodes. Therefore, MPM problem is actually a special case of the influence maximization problem . }

%Define $\Gamma=\{S\subseteq V: c(S)\leq B\}$, that is, $\Gamma$ is the feasible solution set of the MPM problem.

Before presenting the algorithm of MBPM problem, we first introduce another problem, maximization of a monotone submodular function under the cardinality constraint. Let $g: 2^{V}\rightarrow R_{\geq 0}$ be a monotone submodular function. For $u\in V$ and $S\subseteq V$, the marginal gain by adding $v$ to $S$ is denoted as $g_{v}(S)=g(S\cup \{v\})-g(S)$. For the problem $\max_{S\subseteq V, |S|\leq k}{g(S)}$, classical greedy scheme could return $(1-1/e)$-approximation solutions \cite{Nemhauser}. The algorithm always selects the element with largest marginal gain to current selected set until $k$ nodes are chosen. That is, for current selected set $S_{0}$, the algorithm will select $v^{*}=\arg \max_{v\in V\setminus S_{0}}{g_{v}(S_{0})}$ and add it into $S_{0}$. Under cardinality constraint, each node actually has a cost of 1 and greedy scheme always selects the element with the largest marginal gain per unit cost. 

%The constraint $\mathbb{E}[c(S)]\leq B$ in MBPM problem is similar to the cardinality constraint. 

Since the objective of MBPM problem is monotone submodular, inspired by ideas of classical greedy scheme, we could utilize Algorithm \ref{MGA} to solve it. Assume current selected set is $S$. Alg. \ref{MGA} always selects the node $v^{*}$ with largest ratio of marginal gain to $S$ to cost among remaining nodes. If $c(S)+c(v^{*})\leq B$,  $v^{*}$ will be added into $S$. Otherwise, add $v^{*}$ to $S$ with $\frac{B-c(S)}{c(v^{*})}$ probability or break with probability $1-\frac{B-c(S)}{c(v^{*})}$. It could guarantee that the output $S$ satisfies $\mathbb{E}[c(S)]\leq B$. For the node found in the last iteration of our Algorithm, it will be selected with a very low probability if it is far more than the remaining budget.

\begin{algorithm}[htbp]
\caption{Modified Greedy Algorithm} 
\label{MGA}
\hspace*{0.02in} {\bf Input:} 
$G=(V, E)$ and a positive number $B$\\
\hspace*{0.02in} {\bf Output:} 
A seed set $S\subseteq V$.
\begin{algorithmic}[1]
\State $S\leftarrow \emptyset$;
\While{$c(S)< B$}
       \State $v^{*}=\arg \max_{v\in V\setminus S} \frac{P(S\cup \{v\})-P(S)}{c(v)}$;
       \If{$c(S)+c(v^{*})>B$}
               \State break with $1-\frac{B-c(S)}{c(v^{*})}$ probability;
       \EndIf
       \State $S\leftarrow S\cup \{v^{*}\}$
\EndWhile
\State Return $S$;
\end{algorithmic}
\end{algorithm}

Let $v_{1}, \ldots, v_{n}$ be the result sorted by increasing order of activation cost of nodes in $V$. That is, $c(v_{1})\leq c(v_{2})\leq \ldots \leq c(v_{n})$. Denote $p$ as the minimum number satisfying $\sum_{i=1}^{p}{c(v_{i})}\geq B$. Assume $p<n$. Otherwise, we could select all nodes as the seeds. Then we know Algorithm \ref{MGA} would execute at most $p$ iterations.
\begin{theorem}\label{theoratio}
Algorithm \ref{MGA} could achieve a $(1-1/e)$ expected approximation guarantee of MBPM problem. The algorithm requires $O(n^{2})$ function value computation.
\end{theorem}
\begin{IEEEproof}
Since the expected knapsack constraint is somewhat different from the classical knapsack constraint, we think it's necessary to provide the proof for this theorem here. Our proof is inspired by \cite{Nemhauser}. Assume $S^{*}=\{u_{1}, \ldots, u_{t}\}$ is an optimal solution to MBPM problem. Let $S_{r}=\{v_{1},\ldots, v_{r}\}$ be the node set obtained by Algorithm \ref{MGA} after $r$ iterations and $S_{0}=\emptyset$. Assume $v_{r+1}=\arg \max_{v\in (V\setminus S_{r})}\{\frac{P(S_{r}\cup \{v\})-P(S_{r})}{c(v)}\}$. Assume $c(S_{r})\leq B$  and $c(S_{r})+c(v_{r+1})> B$. Let $S_{G}$ be the node set returned by Algorithm \ref{MGA}. Denote $P_{v}(S_{i})=P(S_{i}\cup \{v\})-P(S_{i})$. Since $P(\cdot)$ is monotone  submodular,  for $1\leq i\leq r$, we have
\begin{equation}
 \begin{aligned}
  P(S^{*})&\leq P(S^{*}\cup S_{i})\\
                &=P(S_{i})+P_{u_{1}}(S_{i})+P_{u_{2}}(S_{i}\cup\{u_{1}\})+\ldots\\
                &+P_{u_{t}}(S_{i}\cup\{u_{1},\ldots, u_{t-1}\})\nonumber\\
                &\leq P(S_{i})+P_{u_{1}}(S_{i})+P_{u_{2}}(S_{i})+\ldots+P_{u_{t}}(S_{i})\nonumber\\
         &\leq P(S_{i})+c(u_{1})\cdot \frac{P_{v_{i+1}}(S_{i})}{c(v_{i+1})}+\ldots +c(u_{t})\cdot \frac{P_{v_{i+1}}(S_{i})}{c(v_{i+1})}\\
         &\leq  P(S_{i})+B\cdot \frac{P(S_{i+1})-P(S_{i})}{c(v_{i+1})}.
 \end{aligned}
\end{equation} 
Denote $a_{i}=P(S^{*})-P(S_{i})$. Then we know $a_{i}\leq \frac{B}{c(v_{i+1})}(a_{i}-a_{i+1})$. Thus, 
 \[a_{i+1}\leq (1-\frac{c(v_{i+1})}{B})a_{i}\leq \Pi_{j=1}^{i+1}\left(1-\frac{c(v_{j})}{B}\right)a_{0}\]
 Therefore, \[P(S_{i+1})\geq \left(1-\Pi_{j=1}^{i+1}\left(1-\frac{c(v_{j})}{B}\right)\right)P(S^{*}).\]
 Then we have 
 \begin{equation}
 \begin{aligned}
&\mathbb{E}[ P(S_{G})]= \frac{B-c(S_{r})}{c(v_{r+1})}P(S_{r+1})+\left(1-\frac{B-c(S_{r})}{c(v_{r+1})}\right)P(S_{r})\\
                %&=P(S_{r})+\frac{B-c(S_{r})}{c(v_{r+1})}(P(S_{r+1})-P(S_{r}))\\
               &\geq P(S_{r})+(B-c(S_{r}))\cdot \frac{P(S^{*})-P(S_{r})}{B}\\
               &=\left(1-\frac{c(S_{r})}{B}\right)P(S^{*})+\frac{c(S_{r})}{B}P(S_{r})\\
               &\geq \left(1-\frac{c(S_{r})}{B}\right)P(S^{*})+\frac{c(S_{r})}{B}\left(1-\prod_{j=1}^{r}\left(1-\frac{c(v_{j})}{B}\right)\right)P(S^{*})\nonumber\\
               &=\left(1-\prod_{j=1}^{r}\left(1-\frac{c(v_{j})}{B}\right)\cdot \frac{c(S_{r})}{B}\right)P(S^{*})\nonumber\\
               &= \left(1-\prod_{j=1}^{r}\left(1-\frac{c(v_{j})}{B}\right)\left(1-\left(1-\frac{c(S_{r})}{B}\right)\right)\right)P(S^{*})\nonumber\\
               &\geq 1-\exp\left(-\sum_{j=1}^{r}{\frac{c(v_{j})}{B}}\right)\cdot \exp\left(-\left(1-\frac{c(S_{r})}{B}\right)\right)P(S^{*})\nonumber\\
               &=(1-1/e)P(S^{*})\nonumber\\
 \end{aligned}
\end{equation} 

\end{IEEEproof}
\section{Adaptive Multi-feature Budgeted Profit  Maximization Problem}\label{AMBPM}
In practice, the decision maker may select one seed at a time and then observe the propagation result. He could make choice to select the next seed based on currently observed results. And this strategy is usually called adaptive seed selection strategy. This strategy may bring more advantages and profits since the decision maker could adaptively revise the strategy according to the current situation rather than select all seeds once before the actual propagation process starts. Therefore, it is worth considering whether adaptive selection strategy helps a lot or not. In this section, we will introduce the adaptive MBPM problem and some related definitions.

\subsection{Problem Definition}
In the adaptive MBPM problem, we also choose seeds $S$ from $G=(V, E)$ and observe the propagation process of corresponding seeds $\widehat{S}$ in its multi-level graph $\widehat{G}$ like in the non-adaptive MBPM problem. But under the adaptive setting, seeds are selected one by one and we need observe the diffusion result once a seed $u$ is chosen. Specifically, thresholds for each node $v\in V$ are sampled independently uniformly at random from $[0, 1]$ at first. Then we select one seed at each step. When node $u$ is selected as the next seed, equivalently we infect all of its feature nodes $u_{1}, \ldots, u_{q}$. Then we need to observe states of edges in $\widehat{G}$: observe the propagation result of $u_{i}$ on $G^{i}$ (related edges are live or blocked), $i=1,\ldots, q$. After all $q$ dif\mbox{f}usions on $\widehat{G}$ stop, we could  determine whether nodes in $G$ not buying the product before selecting $u$ would purchase the product or not now, according to the current propagation results on $\widehat{G}$. Then we select next seed and repeat this process until there is no seed budget.

In this adaptive seeding process, after selecting a node $u\in V$ and all feature nodes $u_{1}, \ldots, u_{q}$ of $u$ are infected, we could observe all edges exiting $v_{i}, i=1, \ldots, q$, which can be reached from $u_{i}$ by currently live edges in $G^{i}$. That is, the full-adoption feedback model \cite{Golovin} is considered in this paper.
%Then we could update states of related nodes in $V$ according to states of edges in $\widehat{E}$. 
%If $v$ has been observed to buy the product, then the state of $v$ needs to be updated to its threshold. Otherwise, the state of $v$ remains unknown. 
Our observation so far could be described by the partial realization $\varphi$, a function mapping from currently observed items to their states. For $(u_{i}, v_{i})\in \widehat{E}$, $\varphi((u_{i}, v_{i}))\in \{0, 1, ?\}$ and $\varphi((u_{i}, v_{i}))=1$ (resp. 0) if edge $(u_{i}, v_{i})$ has been observed live (resp. blocked). $\varphi((u_{i}, v_{i}))=?$ if the status of $(u_{i}, v_{i})$ is not known yet.

For any partial realization $\varphi$, define the domain $\text{dom}(\varphi)$ of $\varphi$ as the seed set for the product that have already been picked from $V$. Denote $dx(\varphi)$ as the set of edges in $\widehat{E}$ whose states have been known under $\varphi$.
%Denote $dx(\varphi)$ and $dy(\varphi)$ as the set of edges in $\widehat{E}$ and users in $V$ whose states have been known under $\varphi$, respectively. 
Let $\phi:\widehat{E}\rightarrow \{0,1\}$ be a full realization of $\widehat{G}$. We say a partial realization $\varphi$ is consistent with $\phi$ if they are equal everywhere in the domain of $\varphi$, denoted by $\phi \sim \varphi$. If $\varphi$ and $\varphi'$ are both consistent with some full realization $\phi$, satisfying $\text{dom}(\varphi)\subseteq \text{dom}(\varphi')$, we say $\varphi$ is a subrealization of $\varphi'$, denoted as $\varphi \subseteq \varphi'$.

Let $\pi(\tau)$ be a randomized policy where $\tau$ represents all random source of the randomized policy. Specifically, $\pi(\tau)$ is a function mapping from an already chosen seed set $S\subseteq V$ and a set of partial realizations to $V$, specifying which node to select as the next seed of the product within the budget. Let $S(\pi(\tau), \phi)$ be the set of nodes in $V$ chosen by $\pi(\tau)$ under realization $\phi$. Let $I_{\phi}^{i}(S(\pi(\tau), \phi))$ be the set of nodes in $V^{i}$ accepting feature $i$ when diffusion process of feature nodes of $S(\pi(\tau), \phi)$ on $\phi$ terminates. 
\iffalse
Then for any $u\in V$, we know $u$ would buy the product if and only if 
\[\sum_{\substack{1\leq i \leq q:\\  u_{i}\in I_{\phi}^{i}(S(\pi(\tau),\phi))}}w_{u}^{i}\geq \theta_{u}\]
\fi
The profit obtained by policy $\pi(\tau)$ under realization $\phi$ is defined as:
\[
f\left(S(\pi(\tau), \phi), \phi\right)=\sum_{u\in V}{b(u)}\cdot [\sum_{\substack{1\leq i \leq q:\\  u_{i}\in I_{\phi}^{i}(S(\pi(\tau),\phi))}}{w_{u}^{i}}].
\]
\iffalse
Define 
\[
I_{\phi_{\theta}}(S(\pi(\tau), \phi_{\theta}))=\{u\in V|\sum_{\substack{1\leq i \leq q:\\  u_{i}\in I_{\phi_{\theta}}^{i}(S(\pi(\tau),\phi))}}w_{u}^{i}\geq \theta_{u}\}.
\]
That is, $I_{\phi_{\theta}}(S(\pi(\tau), \phi_{\theta}))$ is the set of nodes that would buy the product under realization $\phi_{\theta}$.
\[
f\left(S(\pi(\tau), \phi_{\theta}), \phi_{\theta}\right)=\sum_{u\in I_{\phi_{\theta}}(S(\pi(\tau), \phi_{\theta}))}{b(u)}.
\]
Let 
\begin{equation}
\begin{aligned}
f\left(S(\pi(\tau), \phi), \phi\right)&=\mathbb{E}_{\theta}\left[f\left(S(\pi(\tau), \phi_{\theta}), \phi_{\theta}\right)\right]\nonumber\\
&=\mathbb{E}_{\theta}[\sum_{u\in I_{\phi_{\theta}}(S(\pi(\tau), \phi_{\theta}))}{b(u)}]\nonumber\\
&=\sum_{u\in V}{b(u)\cdot [\sum_{\substack{1\leq i \leq q:\\  u_{i}\in I_{\phi}^{i}(S(\pi(\tau),\phi))}}w_{u}^{i}}].\nonumber
\end{aligned}
\end{equation}
%Let \[I(S(\pi(\tau), \phi_{\theta}))=\{u\in V| \sum_{\substack{1\leq i \leq q:\\  u_{i}\in I_{\phi_{\theta}}^{i}(S(\pi(\tau),\phi_{\theta}))}}w_{u}^{i}\geq \theta_{u}\}.\] 
\fi
Thus, the expected profit obtained by policy $\pi(\tau)$ can be formulated as:
\[
  f_{\text{avg}}(\pi(\tau))=\mathbb{E}_{\Phi}\left[f\left(S(\pi(\tau), \Phi), \Phi\right)\right].\\
\]
 
\begin{definition}[Adaptive Multi-feature Budgeted Profit  Maximization (AMBPM) Problem]
 Given $G=(V, E)$, assume $q$ pieces of information about $q$ features of a product are spread on $G$ according to the MF-model. The AMBPM problem seeks for a randomized policy to maximize the total expected profit obtained:
 \[
\max_{\pi}{\mathbb{E}_{\tau}[f_{\text{avg}}(\pi(\tau))}]
  \]
  \[
s.t.\ \mathbb{E}_{\tau}[c(S(\pi(\tau), \phi))]\leq B,\ \text{for any realization}\ \phi
  \]
 %adaptively selects a seed set $S\subseteq V$ within the cost budget $B$ to maximize the profit obtained. That is,  seeds for the product are chosen step by step and the next seed is selected within the budget according to current diffusion results. 
\end{definition}

\begin{definition}[Conditional Expected Marginal Profit]
Given a partial realization $\varphi$ and a node $u$, the conditional expected marginal profit of $u$ conditioned on having observed $\varphi$ is defined as:
\[
\Delta(u|\varphi)=\mathbb{E}\left[f(dom(\varphi)\cup\{u\}, \Phi)-f(dom(\varphi), \Phi)|\Phi \sim \varphi\right]
\]
where the expectation is taken over $p(\phi|\varphi)=\mathbb{P}(\Phi=\phi|\Phi\sim \varphi)$.
\end{definition}

\begin{definition}[Adaptive Monotonicity]
A function $f(\cdot, \phi)$ is adaptive monotone with respect to distribution $p(\phi)$, if for all partial realization $\varphi$ with $\Pr[\Phi\sim \varphi]>0$ and all $u\notin \text{dom}(\varphi)$, we have 
\[
\Delta(u|\varphi)\geq 0.
\]
\end{definition}

\begin{definition}[Adaptive Submodularity]
A function $f(\cdot, \phi)$ is adaptive submodular with respect to distribution $p(\phi)$,  if for all partial realizations $\varphi$ and $\varphi'$ satisfying $\varphi\subseteq \varphi'$ and for all $u\notin \text{dom}(\varphi')$, we have 
\[
\Delta(u|\varphi)\geq \Delta(u|\varphi').
\]
\end{definition}
%That is, the conditional expected marginal profit of any fixed node does not increase when more nodes are selected and the edges related to them are observed. 

\begin{theorem}\label{adaptivity}
The objective function $f(\cdot, \phi)$  is adaptive monotone and adaptive submodular.
\end{theorem}
\begin{IEEEproof}
We first show adaptive monotonicity of $f$. Consider a fixed partial realization $\varphi$. For a node $u\notin \text{dom}(\varphi)$, when selecting $u$ as the seed under $\varphi$, if all feature nodes $u_{1}, \ldots, u_{q}$ of $u$ have been infected before $u$ is selected under $\varphi$, then for any realization $\phi\sim \varphi$, we have $f(dom(\varphi)\cup\{u\}, \phi)=f(dom(\varphi), \phi)$. Otherwise, there exists at least one of $u_{1}, \ldots, u_{q}$ not infected before $u$ is selected,  and assume $u_{1}$ is one of the feature node satisfying the condition. Then for any realization $\phi\sim \varphi$, we have $f(dom(\varphi)\cup\{u\}, \phi)-f(dom(\varphi), \phi)\geq b(u)\cdot w_{u}^{1}\geq 0$. Thus, no matter which case happens, for any realization $\phi\sim \varphi$, $f(dom(\varphi)\cup\{u\}, \phi)\geq f(dom(\varphi), \phi)$ always holds. Since $\Delta(u|\varphi)$ is a linear combination of each realization $\phi\sim \varphi$, we know that $\Delta(u|\varphi)\geq 0$.

%then nodes in $\widehat{u}$ will activate some of their out-neighbors, respectively and the probability that $u's$ out-neighbors will purchase the product will increase under any realization $\phi$. But at least the probability that each node will purchase the product under any realization won't decrease. Therefore, for any realization $\phi\sim \varphi$, we have that $f(dom(\varphi)\cup\{u\}, \phi)\geq f(dom(\varphi), \phi)$. Since $\Delta(u|\varphi)$ is a linear combination of each realization, we know that $\Delta(u|\varphi)\geq 0$.

Next we prove the adaptive submodularity of $f$. For any pairs of partial realizations $\varphi, \varphi'$ satisfying $\varphi \subseteq \varphi'$ and any $u\notin \text{dom}(\varphi')$, we have to show $\Delta(u|\varphi)\geq \Delta(u|\varphi')$. Our proof is inspired by the proof technique in \cite{Golovin} and \cite{guo2020k}.

Consider two fixed partial realizations $\varphi, \varphi'$ satisfying $\varphi \subseteq \varphi'$. Assume there are two realizations $\phi$ and $\phi'$ with $\phi \sim \varphi, \phi'\sim \varphi'$, satisfying $\phi((u_{i}, v_{i}))=\phi'((u_{i}, v_{i}))$ for all $(u_{i}, v_{i})\notin dx(\varphi')$. Thus, $\phi$ and $\phi'$ have the same area $\beta=\varphi\cup (\phi' \setminus \varphi')$.

Let $\sigma(\text{dom}(\varphi)\cup \{u\}, \phi)=\cup_{i=1}^{q}{I_{\phi}^{i}(\text{dom}(\varphi)\cup \{u\}, \phi))}$ be the set of infected feature nodes in $\widehat{G}$ when feature nodes of  $\text{dom}(\varphi)\cup \{u\}$ are seeds under the realization $\phi$. Denote $T=\sigma(\text{dom}(\varphi)\cup \{u\}, \phi)$ and $M=\sigma(\text{dom}(\varphi), \phi)$. Let $N=T\setminus M$. Similarly, denote $T'=\sigma(\text{dom}(\varphi')\cup \{u\}, \phi')$ and $M'=\sigma(\text{dom}(\varphi'), \phi')$, and let $N'=T'\setminus M'$.

We first show that $M\subseteq M'$. Fix $w_{i}\in M$. Then there must exist a path $P_{i}$ from some feature node $v_{i}$ of $v\in \text{dom}(\varphi)$ to $w_{i}$. Therefore, edges on path $P_{i}$ are observed to be live by $\varphi$. Since $\phi \sim \varphi, \phi' \sim \varphi'$ and $\varphi \subseteq \varphi'$, edges observed by $\varphi$ have same states in $\phi$ and $\phi'$. That is, each edge on $P_{i}$ is also live under $\phi'$. Since $\varphi\subseteq \varphi'$, it is clear that $v\in \text{dom}(\varphi')$. Therefore, $w_{i}$ will be $i$-accepted when feature nodes of  $\text{dom}(\varphi')$ are seeds in $\widehat{G}$ under realization $\phi'$, i.e., $w_{i}\in M'$.

We next show $N' \subseteq N$. We prove this by contradiction. Fix $v_{j}\in N'$. Assume $v_{j}\notin N$. Since $v_{j}\in N'$ and $M'\cap N' =\emptyset$, we have that $v_{j}\notin M'$. Since we have proven $M\subseteq M'$, it is obvious that $v_{j}\notin M$. As $v_{j}\in N'$, there must exist some path $P_{j}$ from $u_{j}$ to $v_{j}$ in $\phi'$ but at least one edge on path $P_{j}$ is blocked in $\phi$. Assume one such edge is $(s_{j}, t_{j})$. Since the status of edge $(s_{j}, t_{j})$ is different in realization $\phi$ and $\phi'$, and $\phi$ and $\phi'$ have the same area $\beta$, thus $(s_{j}, t_{j})$ must be observed by $\varphi'$ but not by $\varphi$. Since $(s_{j}, t_{j})$ is observed by $\varphi'$, $s_{j}$ must be infected after selecting $\text{dom}(\varphi')$ according to the full-adoption feedback model. That is, $s_{j}$ and the nodes that can be reachable from $s_{j}$ must be infected after we select $\text{dom}(\varphi')$. Therefore,  $s_{j}$ and the nodes that can be reachable from $s_{j}$, including $v_{j}$, will belong to $M'$, a contradiction.

Define
\begin{equation}
 \begin{aligned}
 &\delta(u|\phi, \phi \sim \varphi) =f(\text{dom}(\varphi)\cup\{u\}, \phi)-f(\text{dom}(\varphi), \phi)\\
=&\sum_{v\in V}b(v){\sum_{\substack{1\leq i \leq q:\\  v_{i}\in T}}{w_{v}^{i}}}-\sum_{v\in V}b(v){\sum_{\substack{1\leq i \leq q:\\  v_{i}\in M}}{w_{v}^{i}}}\nonumber\\
=&\sum_{v\in V}b(v){\sum_{\substack{1\leq i \leq q:\\  v_{i}\in (T\setminus M)}}{w_{v}^{i}}}=\sum_{v\in V}b(v){\sum_{\substack{1\leq i \leq q:\\  v_{i}\in N}}{w_{v}^{i}}}
\end{aligned}
\end{equation} 
Since we have shown that $N' \subseteq N$, we could obtain that $\delta(u|\phi, \phi \sim \varphi) \geq \delta(u|\phi', \phi' \sim \varphi')$. Since $\sum_{\phi\sim \beta}{\Pr[\phi|\phi\sim \beta]}=1$, we know

\begin{equation}
 \begin{aligned}
& \Delta(u|\varphi)=\sum_{\phi\sim \varphi}{\Pr[\phi|\phi\sim \varphi]\cdot \delta(u|\phi, \phi \sim \varphi)}\\
 &=\sum_{\phi'\sim \varphi'}{\Pr[\phi'|\phi'\sim \varphi']\sum_{\phi\sim \beta}{\Pr[\phi|\phi\sim \beta]}\cdot \delta(u|\phi, \phi \sim \varphi)}\\
 &\geq \sum_{\phi'\sim \varphi'}{\Pr[\phi'|\phi'\sim \varphi']\sum_{\phi\sim \beta}{\Pr[\phi|\phi\sim \beta]}\cdot \delta(u|\phi', \phi' \sim \varphi')}\\
 &= \sum_{\phi'\sim \varphi'}{\Pr[\phi'|\phi'\sim \varphi']\cdot \delta(u|\phi', \phi' \sim \varphi')}=\Delta(u|\varphi'),\nonumber\\
 \end{aligned}
\end{equation} 
which completes the proof.
\end{IEEEproof}

\section{Algorithm and Theoretical Analysis}\label{algorithm}
Since the objective $f(\cdot, \phi)$ of AMBPM problem is adaptive monotone and adaptive submodular, we could utilize adaptive greedy policy proposed in \cite{Golovin} to solve it. The seed selection rule of adaptive greedy policy is straightforward, i.e.,  always selecting the node with largest ratio of conditional expected marginal profit to cost. However, given a partial realization $\varphi$ and a node $u\notin \text{dom}(\varphi)$, it is difficult to compute the conditional expected marginal profit $\Delta(u|\varphi)$ since there are almost exponential possible realizations $\phi$ with $\phi\sim \varphi$. This section would consider algorithms of AMBPM problem under both the oracle model and noise model.

\iffalse
\textcolor{red}{Before introducing the algorithms in detail, we would like to present some definitions about the policy to make our algorithm more understandable. Imagine the policy running over time, such that when a policy $\pi$ selects a node $u$, it begins to execute $u$, and finishes executing $u$ after $c(u)$ units of time.}

\begin{definition}[Policy truncation with costs]
The $t$-truncation of a policy $\pi$, denoted by $\pi_{[t]}$, is a randomized policy obtained by running $\pi$ for $t$ units of time. If the last selected node $u$ is not finished by time $t$ and it only runs for $0\leq s <c(u)$ time by time $t$, then we will include $u$ in $\pi_{[t]}$ with probability $s/c(u)$. In this case, we could guarantee $\mathbb{E}[c(S(\pi_{[t]}, \phi))]\leq t$ holds for any realization $\phi$.
\end{definition}

\begin{definition}[Policy concatenation]
Given two adaptive policies $\pi_{1}$ and $\pi_{2}$, define the policy concatenation $\pi_{1}\oplus \pi_{2}$ as the policy which runs $\pi_{1}$ first to completion, and then runs $\pi_{2}$ like from a new start, ignoring the information obtained during the running of $\pi_{1}$.
\end{definition}
\fi

\subsection{Adaptive Greedy Algorithm under the Oracle Model}
Under the oracle model, assume conditional expected marginal profit of any node under any partial realization  can be obtained in constant time. Define a randomized adaptive greedy policy $\pi_{ag}(\tau)$. The main idea of adaptive greedy policy to solve this problem can be seen in Algorithm \ref{AAGA}, which is based on the adaptive greedy policy proposed in \cite{Golovin}. Under the current partial realization $\varphi$ and  seed set $S$, the $\pi_{ag}(\tau)$ would select a node $v_{max}$ satisfying $v_{max}:=\arg \max_{v\in V\setminus S}\{\frac{\Delta(v|\varphi)}{c(v)}\}$. If $c(S)+c(v_{max})\leq B$, then $v_{max}$ is the next seed. Otherwise, $\pi_{ag}(\tau)$ would select $v_{max}$ as the next seed with probability $\frac{B-c(S)}{c(v_{max})}$.  After selecting $v_{max}$, we observe the nodes infected by feature nodes of $v_{max}$, denoted by $A(v_{max})$ and update the partial realization $\varphi$ by changing states of edges related to nodes in $A(v_{max})\cup \hat{v}_{max}$ from $?$ to $0$ or $1$.
\iffalse
And we could observe which nodes would buy the product after the propagation of feature nodes of $v_{max}$ terminates and update $\varphi$ by changing these nodes' states from $?$ to their thresholds. \fi
The algorithm repeats the above process, and terminates until $c(S)\geq B$, or terminates with a probability. In this way, we could guarantee $ \mathbb{E}[c(S)]\leq B$. The random source $\tau$ in this adaptive greedy policy indicates whether to contain the node found in the last iteration.

\iffalse
In this paper, instead of computing the conditional expected marginal profit of every node for each iteration, the accelerated adaptive greedy scheme is applied, which could significantly decrease running time in practice. The idea is similar to the accelerated version of classical greedy algorithm in non-adaptive setting \cite{Minoux}. Assume the current partial realization is $\varphi_{i}$ and we would select $v$ as the next seed. Let $\varphi_{j}$ be a partial realization with $\varphi_{j} \subseteq \varphi_{i}$. Assume $u$ has not been activated by the time $v$ is selected. Then there is no need to compute $\Delta(u|\varphi_{i})$ at all. Since if we have known $\frac{\Delta(u|\varphi_{j})}{c(u)}\leq \frac{\Delta(v|\varphi_{i})}{c(v)}$, then we could obtain $\frac{\Delta(u|\varphi_{i})}{c(u)}\leq \frac{\Delta(v|\varphi_{i})}{c(v)}$ due to the adaptive submodularity of $f$.

The algorithm keeps a priority queue to implement the acceleration of the algorithm (lines 6 to 13). At each iteration, it compares value of $\delta_{max}$ with the maximum priority in the current queue. And we only compute the conditional expected marginal profit under current partial realization of the node with larger priority than $\delta_{max}$, where $\delta_{max}$ stores the current maximum ratio of conditional expected marginal profit to cost.
\fi

\begin{algorithm}[htbp]
\caption{Adaptive-Greedy} 
\label{AAGA}
\hspace*{0.02in} {\bf Input:} 
$G=(V, E)$, its multi-level graph $\widehat{G}$ and $B$\\
\hspace*{0.02in} {\bf Output:} 
A seed set $S\subseteq V$ and $f(S, \varphi)$.
\begin{algorithmic}[1]
\State $S\leftarrow \emptyset$; 
\State $\varphi=\{?\}^{\widehat{E}}$; 
\While{$c(S)< B$}
     \State $v_{max}=\arg \max_{v\in V\setminus S}{\Delta(v|\varphi)/c(v)}$;
          \If{$c(S)+c(v_{max})>B$}
             \State break with $1-\frac{B-c(S)}{c(v_{max})}$ probability;
           \EndIf
     %\If{$c(S)+c(v_{max})\leq B$}
           %\State $S \leftarrow S\cup \{v_{max}\}$;
       %\Else
            %\State Select $v_{max}$ into $S$ with probability $\frac{B-c(S)}{c(v_{max})}$;
     % \EndIf
     \State $S \leftarrow S\cup \{v_{max}\}$;
     \State Observe the node set $A(v_{max})$ infected by feature nodes of $v_{max}$, $1\leq i \leq q$;
     %\State Observe the node set $D(v_{max})\subseteq V$ which would buy the product after infecting feature nodes of $v_{max}$;
     \State Update $\varphi$ by updating states of edges related to nodes in $A(v_{max})\cup \hat{v}_{max}$;
     %\State Update $\varphi$ by updating states of nodes in $D(v_{max})$ to their thresholds;
     %\State Update $\widehat{G}$ by removing nodes in $A(v_{max})\cup \hat{v}_{max}$ and their corresponding edges;
\EndWhile
\State Return $S, f(S, \varphi)$
\end{algorithmic}
\end{algorithm}
Since the objective $f(S, \phi)$ of AMBPM problem is adaptive monotone and adaptive submodular, according to the result in \cite{Golovin}, we have the following conclusion.
\begin{theorem}\label{adaptivity}
The adaptive greedy policy shown in Algorithm \ref{AAGA} could obtain a $(1-1/e)$ expected approximation solution of the AMBPM problem. It requires $O(n^{2})$ function value computations.
\end{theorem}

\subsection{Adaptive Greedy Algorithm under the Noise Model}
This section will present algorithms of AMBPM problem under the noise model. The basic seed selection strategy is similar to that in oracle model, but the difference is we will estimate the conditional expected marginal profit of any node under a fixed partial realization, $\Delta(u|\varphi)$, by the reverse influence sampling technique. However, maximizing the estimation of $\Delta(u|\varphi)$ by sampling technique is likely to obtain an extremely worse node with some probability, although the probability is very small. That is, the node $u^{*}$ maximizing the estimation may not be the optimal solution to $\max_{v\in V\setminus S}{\Delta(u|\varphi)/c(u)}$. In this case, the expected approximation ratio in Theorem \ref{adaptivity} is not guaranteed. 

 \subsubsection{Technique}
 
 \begin{definition}[Reverse Reachable (RR) set \cite{Tang14}]
 For any graph realization $\phi\in \Omega$ and $v_{i}\in \widehat{V}$, the RR set for $v_{i}$ is denoted by $R_{\phi}(v_{i})$, which contains all nodes that could reach $v_{i}$ in $\phi$. $v_{i}$ is called the target node of $R_{\phi}(v_{i})$.
 \end{definition}
Intuitively, RR set $R_{\phi}(v_{i})$ of $v_{i}$ contains feature nodes that are likely to infect $v_{i}$ during the propagation. A random RR set is an RR set whose target node $v_{i}$ is selected randomly from $\widehat{V}$. Given a random RR set $R_{\phi}(v_{i})$ and $\widehat{S}\subseteq \widehat{V}$, we say $\widehat{S}$ covers $R_{\phi}(v_{i})$ if $\widehat{S}\cap R_{\phi}(v_{i})\neq \emptyset$. A set with larger expected influence has a higher probability to cover a random RR set. Specifically, given a graph $G=(V, E)$ and a random RR set $R$, the expected influence $\mathbb{E}[I(S)]$ of a set $S$ in $G$ is $\mathbb{E}[I(S)]=|V|\cdot \Pr[S\cap R\neq \emptyset]$ \cite{Tang14}. Therefore, if we could generate a large number of random RR sets, a set with large expected influence would cover a large amount of the generated random RR sets. We will use this idea in our estimation of the conditional expected marginal profit of a node.

Given a partial realization $\varphi$, let $G_{\varphi}=(V_{\varphi}, E_{\varphi})$ be the subgraph induced by the $i$-unaccepted nodes under $\varphi$, $i=1, \ldots, q$. That is, $G_{\varphi}$  is obtained by deleting all of the $i$-accepted feature nodes and their related edges in $\widehat{G}$, $i=1, \ldots, q$. Let $\Omega_{\varphi}$ be the set containing all realizations of $G_{\varphi}$. Denote $W=\sum_{v_{i}\in V_{\varphi}}{b(v)\cdot w_{v}^{i}}$. Assume each node $v_{i}\in V_{\varphi}$ is selected randomly from $V_{\varphi}$ with probability $\frac{b(v)\cdot w_{v}^{i}}{W}$ as the target node of an RR set.

Given a partial realization $\varphi$ and $u\in V$, let $R_{\varphi}$ be a random RR set generated from a realization $\phi \in \Omega_{\varphi}$. Define 
\[
h(u, R_{\varphi})=
\begin{cases} 
1,  & \mbox{if}\ \widehat{u}\cap R_{\varphi}\neq \emptyset \\
0, & \mbox{otherwise}
\end{cases}
\]
By the reverse Breadth First Search algorithm \cite{Moore}, we could produce a large number of  random RR sets $\mathcal{R}(\varphi)=\{R_{1}, R_{2}, \ldots, R_{\alpha}\}$ of $G_{\varphi}$. Define $F_{\mathcal{R}(\varphi)}(u)=\frac{1}{\alpha}\sum_{j=1}^{\alpha}{h(u, R_{j})}$. Denote 
\begin{equation}\label{rhodefinition}
\rho(u|\varphi)=W\cdot F_{\mathcal{R}(\varphi)}(u)=W\cdot \frac{1}{\alpha}\sum_{j=1}^{\alpha}{h(u, R_{j})}.
\end{equation}
Then the following result holds.

\begin{theorem}\label{expection}
Given a node $u\in V$ and a partial realization $\varphi$, we have  $\mathbb{E}\left[\rho(u|\varphi)\right]=\Delta(u|\varphi)$.
\end{theorem}
\begin{IEEEproof}
Given a realization $\phi\in \Omega_{\varphi}$ and a user node $u\in V$, let $I_{\phi}^{i}(u)$ be the feature nodes infected by feature node $u_{i}$ under $\phi$. Then we have  
\begin{equation}
 \begin{aligned}
 \mathbb{E}&\left[\rho(u|\varphi)\right]= \mathbb{E}[W\cdot \frac{1}{\alpha}\sum_{j=1}^{\alpha}{h(u, R_{j})}] = W\cdot \mathbb{E}[\frac{1}{\alpha}\sum_{j=1}^{\alpha}{h(u, R_{j})}]\\
= & W\cdot\sum_{\phi\sim \Omega_{\varphi}}{\Pr[\phi]}\sum_{v_{i}\in V_{\varphi}}{\Pr[v_{i}]\cdot h(u, R_{\varphi}(v_{i}))}\\
=&\sum_{\phi\sim \Omega_{\varphi}}{\Pr[\phi]}\sum_{v_{i}\in V_{\varphi}}b(v)\cdot w_{v}^{i}\cdot h(u, R_{\varphi}(v_{i}))\\
=&\sum_{\phi\sim \Omega_{\varphi}}{\Pr[\phi]}\sum_{v\in V}b(v)\cdot \sum_{\substack{1\leq i \leq q:\\ v_{i}\in I_{\phi}^{i}(u)}}{w_{v}^{i}}\\
=&\sum_{\phi\sim \Omega}{\Pr[\phi|\phi\sim \varphi]}\cdot \sum_{v\in V}{b(v)}\sum_{\substack{1\leq i \leq q:\\ v_{i}\in \left(I_{\phi}^{i}(\text{dom}(\varphi)\cup \{u\})\setminus I_{\phi}^{i}(\text{dom}(\varphi)\right)}}{w_{v}^{i}}\\
=&\sum_{\phi\sim \Omega}{\Pr[\phi|\phi\sim \varphi]}\cdot [f(\text{dom}(\varphi)\cup \{u\}, \phi)-f(\text{dom}(\varphi), \phi)]\\
=&\Delta(v|\varphi)\nonumber
 \end{aligned}
\end{equation} 
\end{IEEEproof}

%According to Theorem \ref{expection}, $\rho(v|\varphi)$ is an unbiased estimate of the expected profit $\Delta(v|\varphi)$. When $\alpha$ is sufficiently large, $\rho(v|\varphi)$ could be convergent to $\Delta(v|\varphi)$. A natural question is: How to determine the value of $\alpha$, the number of random RR sets that we need to generate? We need to make a balance between the accuracy and the time of generating RR sets.

Given a partial realization $\varphi$ and a set of random RR sets $R(\varphi)$ generated from subgraph $G_{\varphi}$,  define $Q_{R(\varphi)}(u)=F_{\mathcal{R}(\varphi)}(u)/c(u)$.
According to Theorem \ref{expection}, we know $\mathbb{E}[W\cdot Q_{R(\varphi)}(u)]=W\cdot \mathbb{E}[Q_{R(\varphi)}(u)]=\Delta(u|\varphi)/c(u)$. Thus, $W\cdot Q_{R(\varphi)}(u)$ is an unbiased estimation of $\Delta(u|\varphi)/c(u)$. When $|R(\varphi)|$ is sufficiently large, $W\cdot Q_{R(\varphi)}(u)$ could be convergent to $\Delta(u|\varphi)/c(u)$. Thus, we could use $W\cdot Q_{R(\varphi)}(u)$ as an estimation for $\Delta(u|\varphi)/c(u)$.

\begin{algorithm}[htbp]
\caption{Sampled-AdapGreedy (SAG)} 
\label{ASG}
\hspace*{0.02in} {\bf Input:} 
A graph $G=(V, E)$, its multi-level graph $\widehat{G}$ and a budget $B\in \mathbb{R}_{+}$, an error parameter $\epsilon$.\\
\hspace*{0.02in} {\bf Output:} 
A seed set $S\subseteq V$ and $f(S, \varphi)$.
\begin{algorithmic}[1]
\State $S=\emptyset$;
\State $\varphi=\{?\}^{\widehat{E}}$; 
\State $G_{\varphi}=G$;
\State $W=\sum_{u\in V}{b(u)}$;
\State $W^{*}=\min_{v_{i}\in \widehat{V}}{b(v)\cdot w_{v}^{i}}$
\While{$c(S)< B$}
     \State $T= V\setminus S$;
     \State $n_{\varphi}=|T|$;
     \State $v^{*} \leftarrow $ Modif\mbox{i}ed-EPIC($G_{\varphi}, T, W, W^{*}, n_{\varphi}, \epsilon$)
     \If{$c(S)+c(v^{*})>B$}
            \State break with $1-\frac{B-c(S)}{c(v^{*})}$ probability;
     \EndIf
     \State $S \leftarrow S\cup \{v^{*}\}$;    
     \State Observe the node set $A(v^{*})$ infected by the feature nodes of $v^{*}$, $1\leq i \leq q$;
      %\State Observe the node set $D(v^{*})\subseteq V$ which would buy the product after selecting $v^{*}$;
     \State Update $\varphi$ by updating states of edges related to nodes in $A(v^{*})\cup \widehat{v^{*}}$;
     %\State Update $\varphi$ by updating states of nodes in $D(v^{*})$ to their threshold;
      \State $W=W-\sum_{u_{i}\in A(v^{*})}{b(u)w_{u}^{i}}-\sum_{v_{i}\in \widehat{v^{*}}\cap G_{\varphi}}b(v^{*})w_{v}^{i}$; 
     \State Update $G_{\varphi}$ by removing nodes in $A(v^{*})\cup \widehat{v^{*}}$ and their corresponding edges;
     \EndWhile
\\
\Return $S$ and $f(S, \varphi)$;
\end{algorithmic}
\end{algorithm}

Algorithm \ref{ASG} show the adaptive greedy policy with the above sampling technique, named Sampled-AdapGreedy. It is denoted by $\pi_{sag}(\tau, \omega)$ where $\omega$ usually represents the random source of sampling. At each iteration, instead of finding a node maximizing $\Delta(u|\varphi)/c(u)$ from currently unselected user nodes, we select a node $v^{*}$ which could maximize $Q_{R(\varphi)}(u)$, which is obtained by Algorithm \ref{MEPIC} \cite{Huang2}. If $c(v^{*})$ is larger than the current remaining budget, then we add $v^{*}$ into the current seed set with $(B-c(v^{*}))/c(v^{*})$ probability. Otherwise, we add $v^{*}$ to the current seed set, observe the corresponding propagation result on $G_{\varphi}$, and update partial realization $\varphi$ and subgraph $G_{\varphi}$. 

 \subsubsection{Theoretical Analysis}
At each iteration of Algorithm \ref{ASG},  it needs use Algorithm \ref{MEPIC} (line 9 of Alg. \ref{ASG}) to obtain a node which could achieve the maximum of function $Q_{R(\varphi)}(u)$. Alg. \ref{MEPIC} is obtained by modifying the EPIC algorithm proposed in \cite{Huang2}. However, there are some difference between EPIC and Modif\mbox{i}ed-EPIC (MEPIC) algorithm: (1) The seed selected at each iteration is one in MEPIC. (2) The target estimation function in MEPIC is $Q_{R(\varphi)}(u)$ instead of $F_{R(\varphi)}(u)$.

\begin{algorithm}[htbp]
\caption{Modif\mbox{i}ed-EPIC (MEPIC) \cite{Huang2}} 
\label{MEPIC}
\hspace*{0.02in} {\bf Input:} 
A graph $G_{\varphi}=(V_{\varphi}, E_{\varphi}), T, W, W^{*}, n_{\varphi}, \epsilon$\\
\hspace*{0.02in} {\bf Output:} 
An approximately optimal node $u\in V\setminus S$.
\begin{algorithmic}[1]
\State $\delta=0.01\cdot \epsilon /W$;
\State $\epsilon'=(\epsilon-\delta\cdot W)/(1-\delta\cdot W)$;
\State $\bar{\epsilon}=\epsilon'/(1-\epsilon')$;
\State $i_{max}=\lceil \log_{2}{\frac{(2+2\bar{\epsilon}/3)\cdot W}{\bar{\epsilon}^{2}}}\rceil+1$ and $a=\ln(\frac{2\cdot i_{max}}{\delta})$;
\State $\theta_{0}=\frac{1}{W^{*}}\left(\ln{\frac{2}{\delta}}+\ln{(n_{\varphi})}\right)$;
\State Generate two sets of random RR sets $\mathcal{R}_{1}(\varphi)$ and $\mathcal{R}_{2}(\varphi)$ of $G_{\varphi}$ with $|\mathcal{R}_{1}(\varphi)|=|\mathcal{R}_{2}(\varphi)|=\theta_{0}$;
\For{$i=1$ to $i_{max}$}
     \State $v^{*}=\arg \max_{v\in T}Q_{R_{1}(\varphi)}(v)$;
     \State $Q^{u}(v_{max})\leftarrow Q_{R_{1}(\varphi)}(v^{*})$;
     \State $Q^{l}(v^{*})\leftarrow \left(\sqrt{Q_{R_{2}(\varphi)}(v^{*})+\frac{2a}{9 |R_{2}(\varphi)|}}-\sqrt{\frac{a}{2 |R_{2}(\varphi)|}}\right)^{2} -\frac{a}{18\cdot |R_{2}(\varphi)|}$;
     \If{$\frac{Q^{l}(v^{*})}{Q^{u}(v_{max})}\geq 1-\epsilon'$ or $i=i_{max}$}
           %\Return $v^{*}$;
           \State \textbf{return}\ $v^{*}$;
     \EndIf
     \State Double the sizes of $R_{1}(\varphi)$ and $R_{2}(\varphi)$ with new random RR sets;
\EndFor
\end{algorithmic}
\end{algorithm}

At each iteration of Algorithm \ref{ASG}, denote the current partial realization as $\varphi$. We could obtain its corresponding induced subgraph $G_{\varphi}$. Alg. \ref{MEPIC} first initializes some parameters and then generates two same size sets of random RR sets of $G_{\varphi}$, $\mathcal{R}_{1}(\varphi)$ and $\mathcal{R}_{2}(\varphi)$. At each iteration, it chooses a node $v^{*}$ maximizing $Q_{R_{1}(\varphi)}(\cdot)$, which can be achieved in polynomial time. Assume $v_{max}=\arg \max_{v\in T}{\Delta(v|\varphi)/c(v)}$.Then $Q^{u}(v_{max})=Q_{R_{1}(\varphi)}(v^{*})\geq Q_{R_{1}(\varphi)}(v_{max})$. That is, $Q^{u}(v_{max})$ is an upper bound of $Q_{R_{1}(\varphi)}(v_{max})$. And $W\cdot Q^{l}(v^{*})$ is an accurate lower bound of  $\Delta(v^{*}|\varphi)/c(v^{*})$ with high probability. Then MEPIC checks whether the stopping condition (line 11) is satisfied. If satisfied, it returns $v^{*}$ as output. Otherwise, it doubles the size of $\mathcal{R}_{1}(\varphi)$ and $\mathcal{R}_{2}(\varphi)$, and repeats the above process.

 %\textcolor{blue}{A natural question is: How to determine the value of $|R(\varphi)|$, the number of random RR sets that we need to generate? We need to make a balance between the accuracy and the time of generating RR sets.}

\begin{lemma}\label{imlemma}
Given a partial realization $\varphi$ and its corresponding induced subgraph $G_{\varphi}=(V_{\varphi}, E_{\varphi})$, denote by $T$ the set of current unselected nodes in $V$. Then MEPIC algorithm could return a user node $v^{*}$ satisfying that 
\[
\mathbb{E}_{\tau}\left[\frac{\Delta(v^{*}|\varphi)}{c(v^{*})}\right]\geq (1-\epsilon) \cdot \max_{v\in T}\left\{\frac{\Delta(v|\varphi)}{c(v)}\right\},
\]
within $O((|V_{\varphi}|+|E_{\varphi}|)(\log(|T|)+\log{\frac{1}{\epsilon}})/\epsilon^{2})$ expected time.
\end{lemma}
\begin{IEEEproof}
Given a partial realization $\varphi$ and its corresponding induced subgraph $G_{\varphi}=(V_{\varphi}, E_{\varphi})$, the target function $Q_{R_{1}(\varphi)}(v)=F_{R_{1}(\varphi)}(u)/c(u)$ is a weighted coverage function on $R_{1}(\varphi)$, where the weight for each $u\in (V\setminus \text{dom}(\varphi))$ is $1/c(u)$. Since $Q_{R_{1}(\varphi)}(v)$ is a monotone submodular function and maximizing a monotone submodular weighted coverage function can be solved in polynomial time, thus the node $v^{*}=\arg \max_{v\in T}Q_{R_{1}(\varphi)}(v)$ can be obtained in polynomial time. Also, the expected approximation guarantee can be obtained accordingly from results of EPIC algorithm in \cite{Huang2}.
\end{IEEEproof}

Recall that $p$ is the minimum number satisfying $\sum_{i=1}^{p}{c(v_{i})}\geq B$. Then we know Algorithm \ref{ASG} would execute at most $p$ iterations. Now, we could give the approximation guarantee of our AG algorithm.
\begin{theorem}\label{appratio}
Given $\epsilon\in(0, 1)$, the sampled adaptive greedy policy $\pi_{sag}(\tau, \omega)$ (Algorithm \ref{ASG}) could achieve a $(1-e^{-(1-\epsilon)})$ expected approximation ratio within $O(pq\cdot(n+m)(\log(n+\log{\frac{1}{\epsilon}})/\epsilon^{2})$ expected time. That is, for any realization $\phi$ and any policy $\pi(\tau)$ satisfying $\mathbb{E}_{\tau}[c(S(\pi(\tau), \phi))]\leq B$, we have 
\[
\mathbb{E}_{\tau}[\mathbb{E}_{\omega}[f_{avg}(\pi_{sag}(\tau, \omega))]]\geq (1-e^{-(1-\epsilon)})\cdot \mathbb{E}_{\tau}[f_{avg}(\pi(\tau))]
\]
\end{theorem}
\begin{IEEEproof}
According to Lemma \ref{imlemma}, the node selected in each iteration of Algorithm \ref{ASG} satisfies $(1-\epsilon)$ expected approximation. Since Algorithm \ref{ASG} would execute no more than $p$ iterations, thus the total expected error of all iterations is $(1/p)\cdot \sum_{i=1}^{p}{\epsilon}=\epsilon$. According to the Theorem \ref{theoratio} and Lemma \ref{imlemma}, Theorem \ref{appratio} holds by inferring from Theorem 6 in  \cite{Huang2}.
\end{IEEEproof}

\section{Experiments}\label{Experiments}
We verify efficiencies of our proposed policy by comparing the running time and its obtained profit with other algorithms. We run experiments on a Linux machine with an Intel Xeon 3.5GHz CPU and 32GB RAM. For each dataset, 30 possible realizations are produced randomly and the average performance of each algorithm is reported.

\subsection{Experimental Setup}
\textbf{Datasets.} Five real-world social network datasets are used in this paper and detailed statistics are shown in table \ref{Dataset characteristics}. Epinions dataset could be found in \cite{SNAP} and all other datasets are from \cite{Rossi}. According to the structure of MF-model, the number of nodes in multi-level graph is different from these basic information, which is also determined by the number of features. For the undirected graph, we replace each edge with two reversed directed edges.
\begin{table}[hptb]\centering\small
\setlength{\abovecaptionskip}{-0.1cm}
\setlength{\belowcaptionskip}{0.2cm}
\caption{Dataset characteristics}\label{Dataset characteristics}
\begin{tabular}{|p{1.5cm}<{\centering}|p{0.8cm}<{\centering}|p{0.8cm}<{\centering}|p{1.2cm}<{\centering}|p{2.1cm}<{\centering}|}
\hline
\textbf{Dataset} & \textbf{\emph{n}} & \textbf{\emph{m}} & \textbf{Type} & \textbf{Average degree}\\
\hline
\emph{Twitter} & 0.8k & 1k & directed & 2 \\
\hline
\emph{Wiki} & 0.9k & 3k & directed & 6\\
\hline
\emph{Hamsterster} & 2.4k & 16.6k & undirected & 13\\
\hline
\emph{DBLP} & 12.6k & 49.7k & undirected & 7.9\\
\hline
\emph{HepPh} & 12k & 118k & undirected & 19\\
\hline
\emph{Epinions} & 75.9k & 508.8k & directed & 13\\
\hline
\end{tabular}
\label{tab:dataset}
\end{table}

\textbf{Propagation Model and Parameters.} We use the MF-model as diffusion model in experiments and for each edge $e=(u, v)\in E$, set $p_{u, v}^{i}=1/|N^{in}(v)|, i =1,\ldots, q$, where $N^{in}(v)$ is the set of in-neighbors of $v$. This setting is widely used in prior works \cite{Tang14, Goyal, Jung}. For $u\in V$, the weight vector of $u$ is generated randomly from $(0, 1]^{q}$ such that the sum of weights of all features for $u$ is 1. Also, we generate random numbers from $(0, 1]$ as the cost and profit of each node. For each dataset, we vary budget $B$ such that $B\in \{0, 10, 20, 30, 40, 50\}$.

We conduct two groups of experiments to test the time efficiency and performance of our proposed policy, respectively. The first group of experiments is performed to verify the time efficiency of adaptive greedy policy (Algorithm \ref{AAGA}) and sampled adaptive greedy policy (Algorithm \ref{ASG}). We compare running time and obtained profit of adaptive greedy policy and sampled adaptive greedy policy with their non-adaptive corresponding algorithms, with different implementations. \\
(1) Modified greedy algorithm sampled by Monte Carlo (MGMC): Shown as Algorithm \ref{MGA} and the profit $P(S)$ of any node set $S$ is estimated by Monte Carlo method. \\
(2) Modified greedy algorithm sampled by reverse influence sampling (MGRIS): Shown as Algorithm \ref{MGA} and the profit $P(S)$ of any node set $S$ is estimated by reverse influence sampling method. Let $Q=\sum_{u\in V}{b(u)}$. Each feature node $v_{i}$ in multi-level graph $\widehat{G}$ is selected as a target node of a RR set with $b(v)\cdot w_{v}^{i}/Q$ probability. Let $\mathcal{R}=\{R_{1}, \ldots, R_{\lambda}\}$ be a set of random RR sets generated from $\widehat{G}$. Then it can be proved $K_{\mathcal{R}}(S)=Q\cdot F_{\mathcal{R}}(S)$ is an unbiased estimation of $P(S)$. According to Chernoff Bounds \cite{Motwani}, if $\lambda\geq \frac{(2+\eta)Q}{\eta^{2}P(S)}\cdot \ln{\frac{1}{\delta'}}$, then for any node set $S$ with $c(S)\leq B$, we have $\Pr[|P(S)-K_{\mathcal{R}}(S)|>\eta\cdot P(S)]<\delta'$. Let $p^{*}$ be the minimum number such that $\sum_{j=p^{*}}^{n}c(v_{j})\leq B$ and $Q^{*}=\sum_{j=p^{*}}^{n}b(v_{j})$. By setting $\lambda =\frac{(2+\eta)Q}{\eta^{2}Q^{*}}\cdot \ln{\frac{1}{\delta'}}$, we could guarantee  $\lambda\geq \frac{(2+\eta)Q}{\eta^{2}P(S)}\cdot \ln{\frac{1}{\delta'}}$. Here we set $\eta=\delta'=0.1$.\\
(3) Adaptive greedy policy (AG): Shown as Algorithm \ref{AAGA} and the conditional expected marginal profit of a node $u$ under any partial realization, $\Delta(u|\varphi)$, is estimated by Monte Carlo method.\\
(4) Sampled adaptive greedy policy (ASG): Shown as Algorithm \ref{ASG} and set the error parameter $\epsilon=0.5$.

The second group of experiments is to compare the performance of our SAG policy with three heuristic adaptive policies: Adaptive Random (AR), Adaptive Max-degree (AMD) and Adaptive Max-profit (AMP).\\
(1) AR is the adaptive version of the simple random algorithm. It uniformly selects currently unselected nodes in $V$ as the next seed. \\
(2) AMD picks the node with maximum out-degree from currently unselected nodes in $V$ as the next seed. \\
(3) Given the partial realization $\varphi$ and currently selected seed set $S$, AMP selects the node $u^{*}$ satisfying $u^{*}\in \arg \max_{u\in (V\setminus S)}{\Delta(u|\varphi)}$ and estimates $\Delta(u|\varphi)$ by reverse influence sampling technique. According to Theorem \ref{expection}, we know $\rho(u|\varphi)$ is an unbiased estimation of $\Delta(u|\varphi)$. According to Chernoff Bounds \cite{Motwani}, if $\alpha\geq \frac{(2+\hat{\epsilon})W}{\hat{\epsilon}^{2}\Delta(u|\varphi)}\cdot \ln{\frac{1}{\delta'}}$, then we have $\Pr[|\rho(u|\varphi)-\Delta(u|\varphi)|>\hat{\epsilon}\cdot \Delta(u|\varphi)]<\delta'$. Let $W^{*}=\min_{v_{i}\in \widehat{V}}{b(v)\cdot w_{v}^{i}}$. By setting $\alpha =\frac{(2+\hat{\epsilon})W}{\hat{\epsilon}^{2}W^{*}}\cdot \ln{\frac{1}{\delta'}}$, we could guarantee  $\alpha\geq \frac{(2+\hat{\epsilon})W}{\hat{\epsilon}^{2}\Delta(u|\varphi)}\cdot \ln{\frac{1}{\delta'}}$. Here we set $\hat{\epsilon}=\delta'=0.1$.\\

\subsection{Experimental Results}
\subsubsection{Results of first group of experiment}
\begin{figure}[htbp]
\centerline{\includegraphics[scale=0.0483]{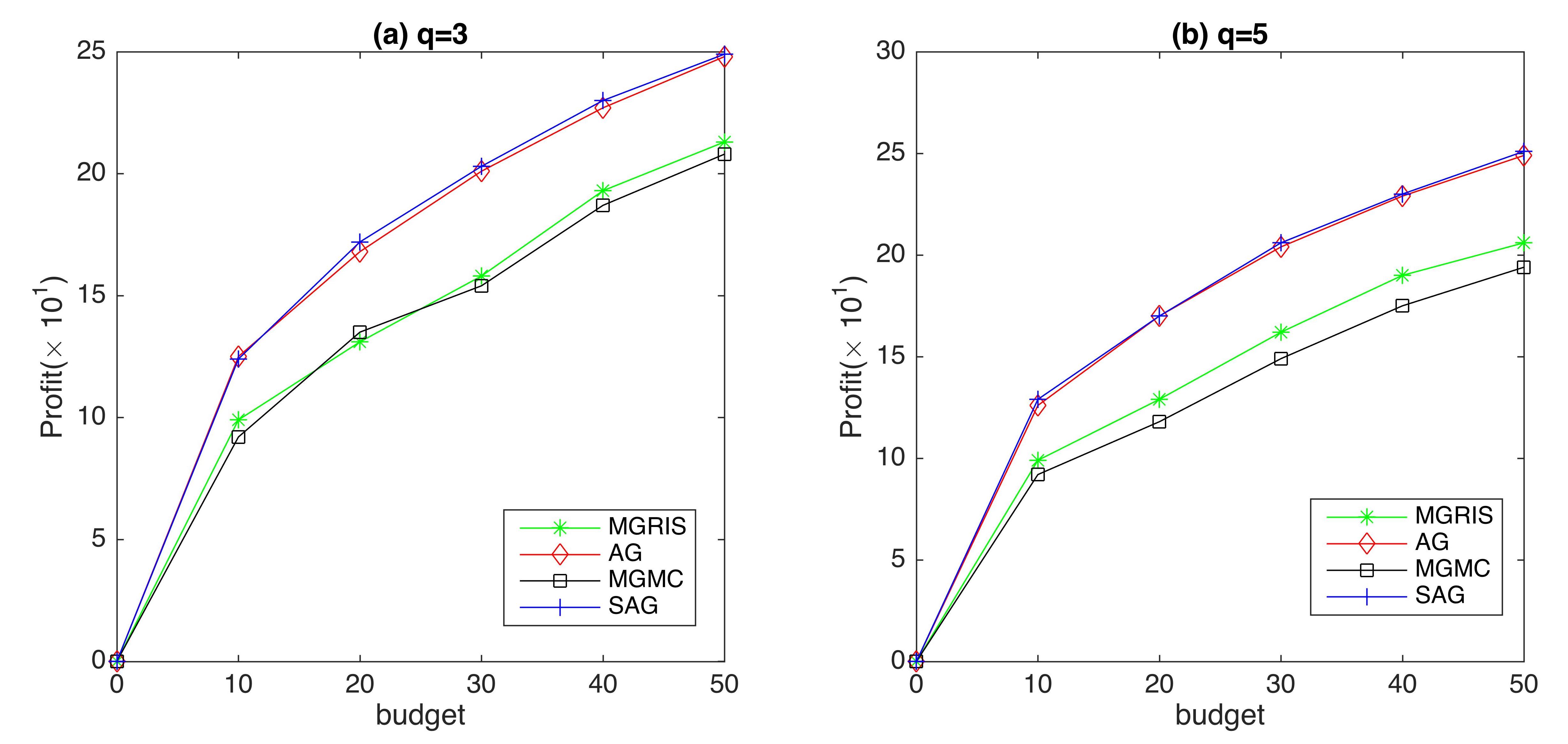}}
  \caption{Profit VS budget on Twitter.}
\label{nonadap-profit1}
\end{figure}

\begin{figure}[htbp]
\centerline{\includegraphics[scale=0.0483]{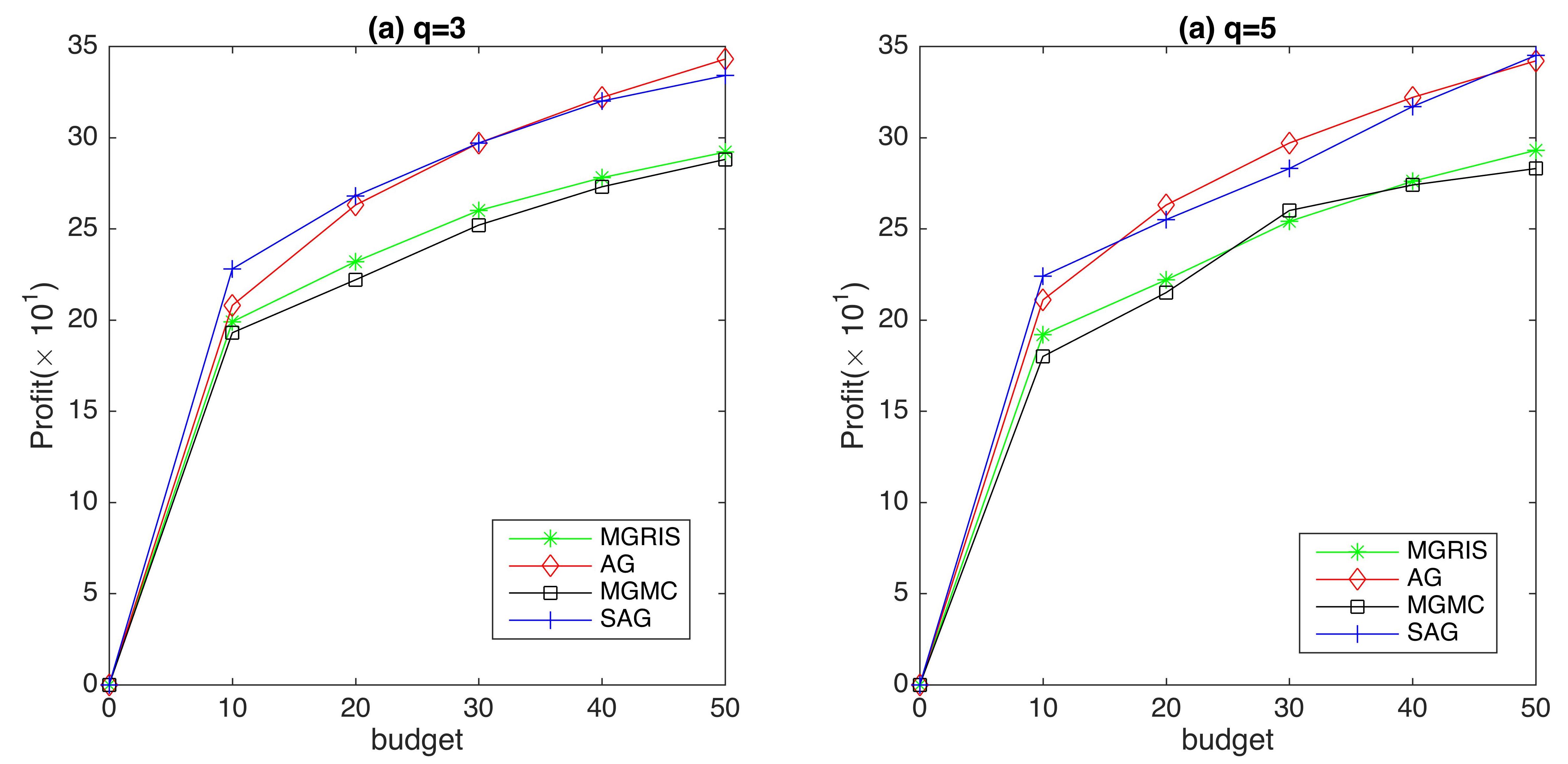}}
  \caption{Profit VS budget on Wiki.}
\label{nonadap-profit2}
\end{figure}

Fig. \ref{nonadap-profit1} and Fig. \ref{nonadap-profit2} present results of the first group of experiments on Twitter and Wiki datasets. Fig. \ref{nonadap-profit1} and Fig. \ref{nonadap-profit2} present profits obtained by our proposed adaptive greedy policy (AG and SAG) with their non-adaptive versions (MGMC and MGRIS). We implement the experiments under two values of $q$, 3 and 5. Since AG policy and MGMC algorithm are implemented by Monte Carlo method and they are time-consuming, thus we only conduct the first group of experiments on two small datasets. Here the number of Monte Carlo simulations is set to $500$.  The results show that AG and SAG policy are always evidently superior than MGMC and MGRIS algorithm with respect to the obtained profit, which shows the benefits of adaptive policies. The profits obtained by AG and SAG policies are very close,  which indicates effectiveness of our sampling technique. The profits obtained by MGRIS and MGMC algorithms are close at most time, but in some cases, profits of MGMC are smaller than those of MGRIS. This may be because the number of Monte Carlo simulations is not enough.


Table \ref{Twitter} presents the running time of our proposed AG and SAG policies with MGMC and MGRIS on Twitter and Wiki datasets under budget 10, 30 and 50, respectively. To compare running time of different strategies fairly, parallel computing is not used here. We can see that MGRIS is fastest among all of the four algorithms since it only need to generate a set of random RR sets once and choose seeds once. Our SAG policy is the second fastest strategy and faster than AG and MGMC. AG is much faster than MGMC and this may be because the induced subgraph of partial realization becomes smaller and smaller. 

\begin{table}[htbp]\centering\small
\setlength{\abovecaptionskip}{-0.1cm}
\setlength{\belowcaptionskip}{0.2cm}
\caption{Running time VS budget on Twitter and Wiki}\label{Twitter}
\begin{tabular}{|p{1.3cm}<{\centering}|p{0.7cm}<{\centering}|p{0.7cm}<{\centering}|p{0.7cm}<{\centering}|p{0.7cm}<{\centering}|p{0.7cm}<{\centering}|p{0.86cm}<{\centering}|}
\hline
\multicolumn{7}{|c|}{\textbf{Twitter}}\\
\hline
 \textcolor{white}{1}&\multicolumn{3}{c|}{\bm{$q=3$}}&\multicolumn{3}{c|}{\bm{$q=5$}}\\
\hline
\textbf{Algorithm} & \textbf{10} & \textbf{30} & \textbf{50}& \textbf{10} & \textbf{30} & \textbf{50}\\
\hline
\emph{MGRIS(s)} & 5.36 & 20.64 &39.57& 9.1& 31.54 &61.3\\
\hline
\emph{MGMC(h)} & 0.7 & 3.01  &5.46& 0.97 & 4.35  &7.57\\
\hline
\emph{AG(s)} & 175.23  & 419.59 &667.5& 271.48& 677.69  &1012.19\\
\hline
\emph{SAG(s)} & 42.64& 137.58  &231.2& 59.84 & 186.75  &313.62\\
\hline
\multicolumn{7}{|c|}{\textbf{Wiki}}\\
\hline
 \textcolor{white}{1}&\multicolumn{3}{c|}{\bm{$q=3$}}&\multicolumn{3}{c|}{\bm{$q=5$}}\\
\hline
\textbf{Algorithm} & \textbf{10} & \textbf{30} & \textbf{50}& \textbf{10} & \textbf{30} & \textbf{50}\\
\hline
\emph{MGRIS(s)} & 7.31 & 25.82 &46.75& 10.37 & 39.03 &68.52\\
\hline
\emph{MGMC(h)} & 3.56& 11.41  &19.01& 6.67& 22.2  &36.92\\
\hline
\emph{AG(s)} & 186.81  & 445.79  &690.42& 268.85 & 639.21  &970.45\\
\hline
\emph{SAG(s)} & 39.29  & 143.84  &239.65& 67.55& 190.42  &323.28\\
\hline
\end{tabular}
\label{tab:dataset}
\end{table}

\subsubsection{Results of second group of experiment}

\begin{figure}[htbp]
\centerline{\includegraphics[scale=0.0245]{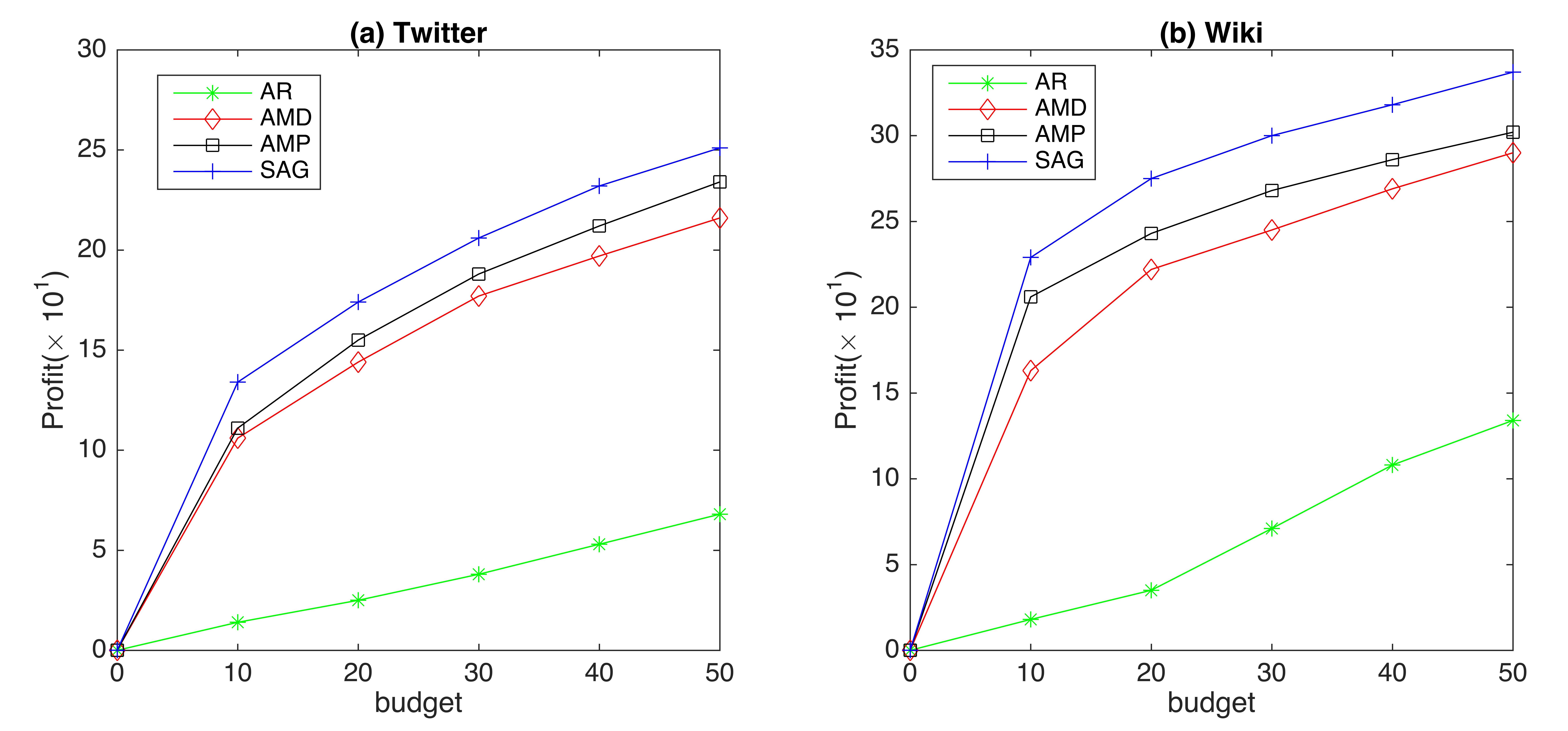}}
  \caption{Profit VS budget on Twitter and Wiki.}
\label{twitter_wiki}
\end{figure}

\begin{figure}[htbp]
\centerline{\includegraphics[scale=0.05]{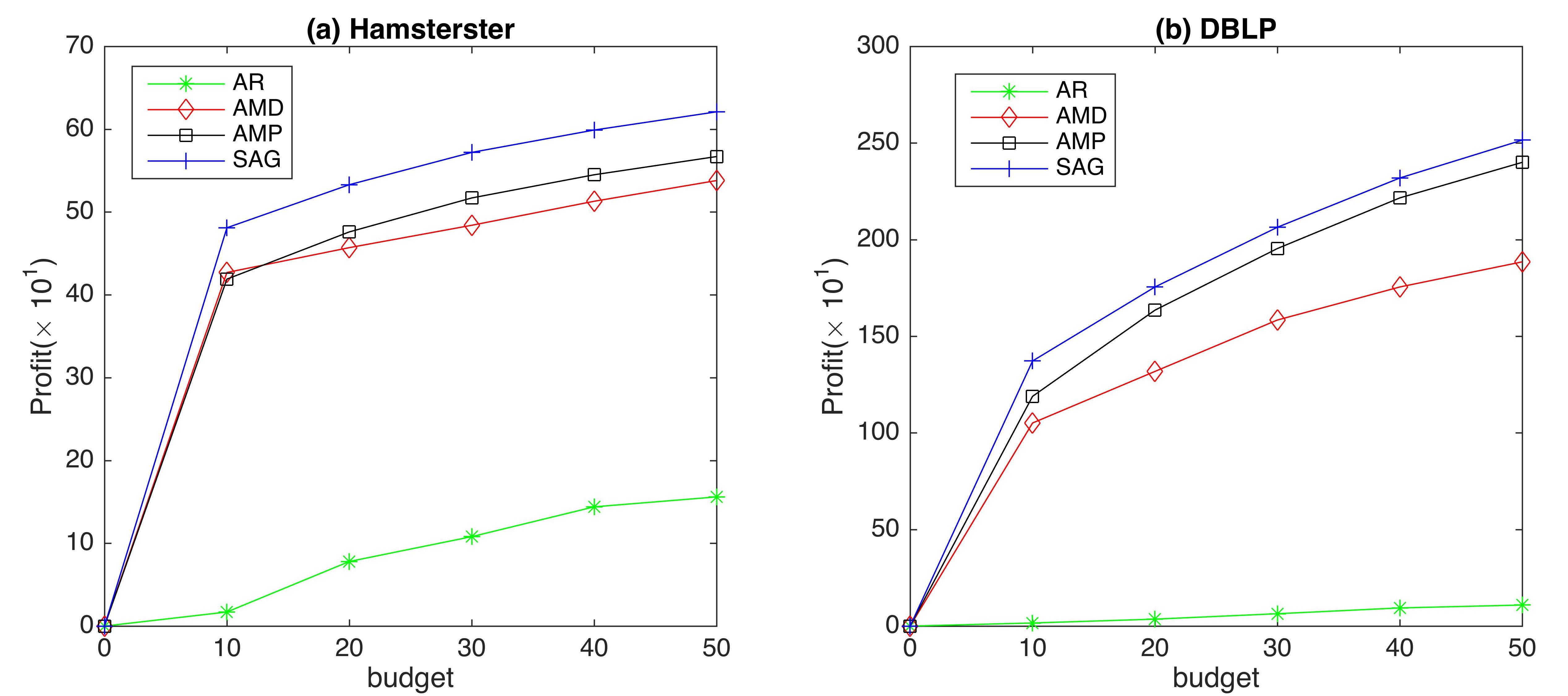}}
  \caption{Profit VS budget on Hamsterster and DBLP.}
\label{hamster_dblp}
\end{figure}

\begin{figure}[htbp]
\centerline{\includegraphics[scale=0.05]{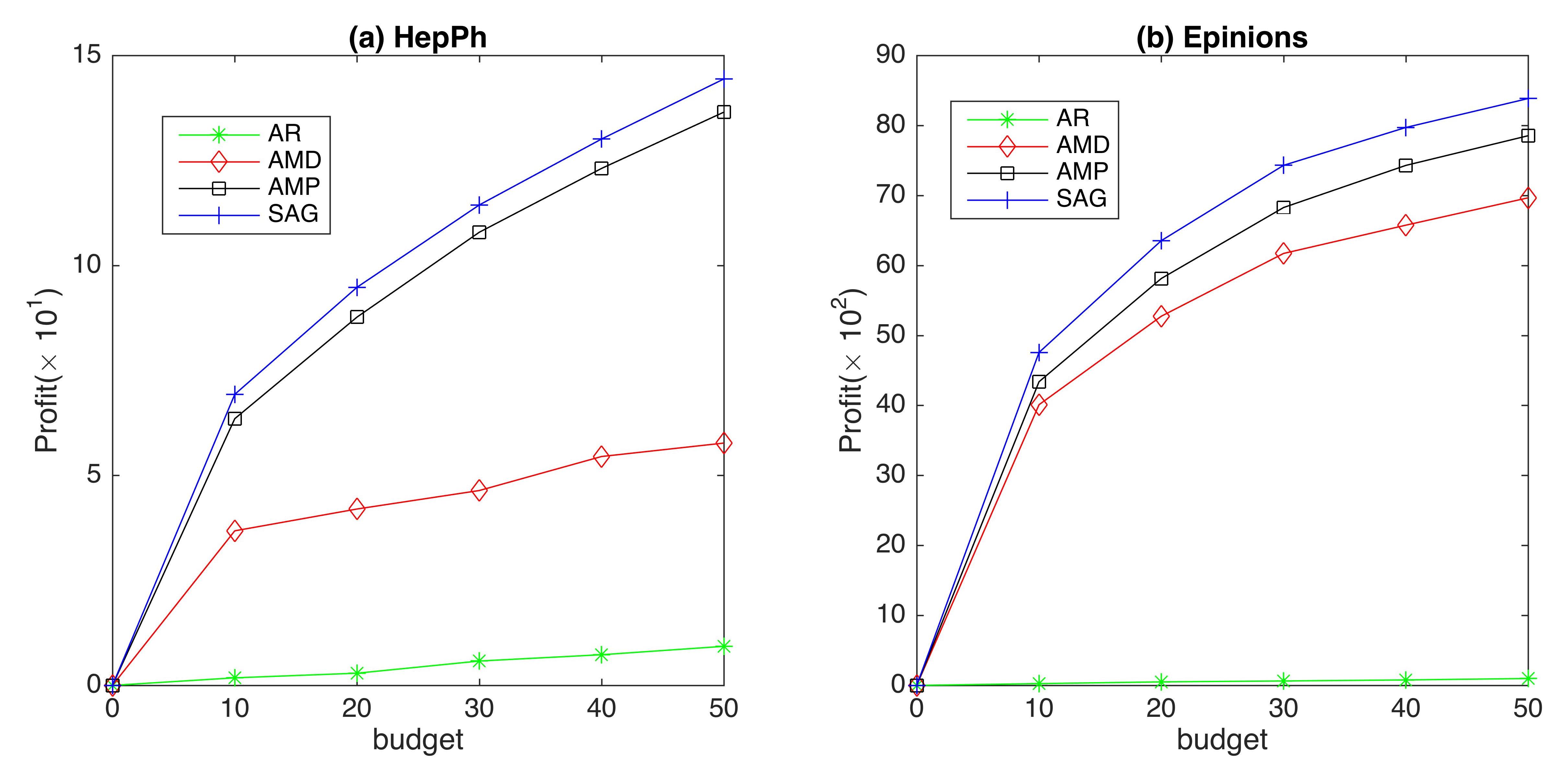}}
  \caption{Profit VS budget on HepPh and Epinions.}
\label{hepph_epinion}
\end{figure}

Fig. \ref{twitter_wiki} to Fig. \ref{hepph_epinion} present the performance of our SAG policy and other three heuristic adaptive policies on all of the six listed datasets. In this group of experiments, the value of $q$ is set to 3. We can see that the profits obtained by any policies increase with the value of the budget. And profits obtained by our SAG policy are always higher than those obtained by other three heuristic adaptive policies no matter on which one of the six datasets. Among the three heuristic adaptive policies, adaptive max-profit (AMP) policy performs better than AR and AMD policies at most time. This is intuitive since AMP considers the profit not just the degree and a node with large degree may not bring many profits. And our SAG policy usually can obtain about $10\%$ higher profits than AMP policy, which also indicates the effectiveness of our  SAG policy. But the results are not so stable and this may be due to the different graph structures and other features of different datasets.

\section{Conclusion}\label{Conclusion}

This work proposes a novel problem, multi-feature budgeted pro\mbox{f}it maximization problem (MBPM), which asks for a seed set with expected cost no more than the budget to make expected pro\mbox{f}it as large as possible. We mainly consider the adaptive MBPM problem, where the seeds are selected iteratively and next seed is chosen based on the current diffusion result. We study the adaptive MBPM problem under two models, oracle model and noise model. Speci\mbox{f}ically, a $(1-1/e)$ expected approximation policy is proposed in the oracle model. Under the noise model, we compute conditional expected marginal pro\mbox{f}it of a node under a partial realization by reverse in\mbox{f}luence sampling technique and propose an e\mbox{ff}icient algorithm, which could achieve a $(1-e^{-(1-\epsilon)})$ expected approximation ratio, where $0<\epsilon<1$. To evaluate the performance of our algorithms, extensive experiments are done on six realistic datasets with the comparison of our proposed policies to their corresponding non-adaptive algorithms and some heuristic adaptive policies.

\section{Acknowledgement}

This work is supported in part by NSF under grants 1747818 and 1907472.

\bibliographystyle{IEEEtran}
\bibliography{IEEEabrv,sample}

\iffalse

\fi

\begin{IEEEbiography}[{\includegraphics[width=1in,height=1.25in,clip,keepaspectratio]{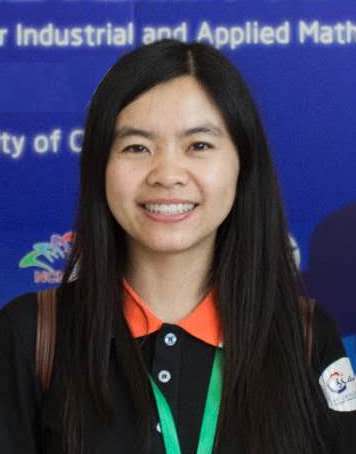}}]{Tiantian Chen}
is a Ph.D. candidate in the Department of Computer Science, The University of Texas at Dallas. She received her B.S. degree in Mathematics and Applied Mathematics, and M.S. degree in Operational Research and Cybernetics from Ocean University of China in 2016 and 2019, respectively. Her research focuses on design and analysis of approximation algorithms and social networks.
 \end{IEEEbiography}
\vspace{-0.5cm}
\begin{IEEEbiography}[{\includegraphics[width=1in,height=1.2in,clip,keepaspectratio]{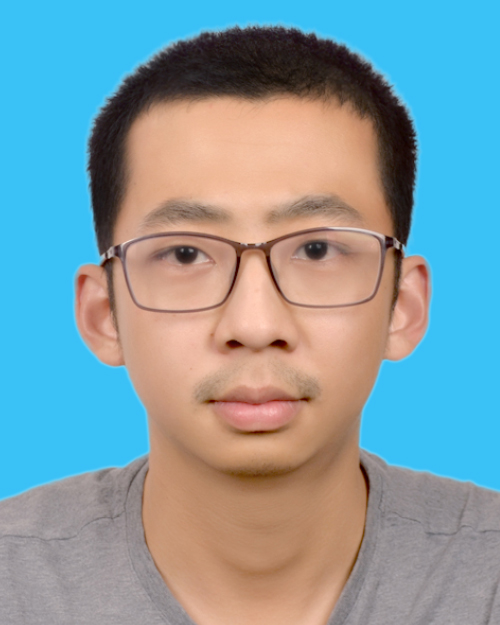}}]{Jianxiong Guo}
 received his Ph.D. degree from the Department of Computer Science, University of Texas at Dallas, Richardson, TX, USA, in 2021, and his B.E. degree from the School of Chemistry and Chemical Engineering, South China University of Technology, Guangzhou, Guangdong, China, in 2015. He is currently an Assistant Professor with the BNU-UIC Institute of Artificial Intelligence and Future Networks, Beijing Normal University at Zhuhai, and also with the Guangdong Key Lab of AI and Multi-Modal Data Processing, BNU-HKBU United International College, Zhuhai, Guangdong, China. His research interests include social networks, algorithm design, data mining, IoT application, blockchain, and combinatorial optimization.
\end{IEEEbiography}
\vspace{-0.5cm}
\begin{IEEEbiography}[{\includegraphics[width=1in,height=1.25in,clip,keepaspectratio]{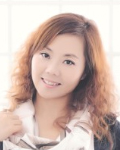}}]{Weili Wu}
 received the Ph.D. and M.S. degrees from the Department of Computer Science, University of Minnesota, Minneapolis, MN, USA, in 2002 and 1998, respectively. She is currently a Full Professor with the Department of Computer Science, The University of Texas at Dallas, Richardson, TX, USA. Her research mainly deals in the general research area of data communication and data management. Her research focuses on the design and analysis of algorithms for optimization problems that occur in wireless networking environments and various database systems.
\end{IEEEbiography}

\end{document}